\documentclass[12pt]{article}
\makeatletter
\newcommand{\singlespacing}{\let\CS=\@currsize\renewcommand{\baselinestretch}{1}\tiny\CS}
\usepackage{epsfig}
\usepackage{graphicx}
\usepackage{verbatim}   % useful for program listings
\usepackage{color}      % use if color is used in text
\usepackage{amsfonts}
\usepackage{amsmath}
\usepackage{amssymb}
\usepackage{tikz}
\usepackage{makeidx}
\usepackage[left=2cm,right=2cm,top=2cm,bottom=2cm]{geometry}
\usetikzlibrary{patterns}
\usepackage{float}
\usepackage{booktabs}
\usepackage{multirow}
\usepackage{caption}
\usepackage{subcaption}
\usepackage{array}% http://ctan.org/pkg/array
\usepackage{booktabs}
\usepackage{multirow}
\newcolumntype{d}{D{.}{.}{2.5}}
\newcolumntype{C}{>{\centering}p}
\newtheorem{theorem}{Theorem}[section]

\newtheorem{corollary}[theorem]{Corollary}

\begin{document}
\baselineskip=24pt
%\singlespacing
%\doublespacing
\parskip = 10pt
\def \qed {\hfill \vrule height7pt width 5pt depth 0pt}
\newcommand{\ve}[1]{\mbox{\boldmath$#1$}}
\newcommand{\IR}{\mbox{$I\!\!R$}}
\newcommand{\1}{\Rightarrow}
\newcommand{\bs}{\baselineskip}
\newcommand{\esp}{\end{sloppypar}}
\newcommand{\be}{\begin{equation}}
\newcommand{\ee}{\end{equation}}
\newcommand{\beanno}{\begin{eqnarray*}}
\newcommand{\inp}[2]{\left( {#1} ,\,{#2} \right)}
\newcommand{\eeanno}{\end{eqnarray*}}
\newcommand{\bea}{\begin{eqnarray}}
\newcommand{\eea}{\end{eqnarray}}
\newcommand{\ba}{\begin{array}}
\newcommand{\ea}{\end{array}}
\newcommand{\nno}{\nonumber}
\newcommand{\dou}{\partial}
\newcommand{\bc}{\begin{center}}
\newcommand{\ec}{\end{center}}
\newcommand{\2}{\subseteq}
\newcommand{\cl}{\centerline}
\newcommand{\ds}{\displaystyle}
\newcommand{\what}{\widehat}
\def\refhg{\hangindent=20pt\hangafter=1}
\def\refmark{\par\vskip 2.50mm\noindent\refhg}

\title{\sc Inference on Weibull Parameters Under a Balanced Two Sample Type-II Progressive Censoring Scheme 
 }
\author{\sc Shuvashree Mondal$^1$, Debasis Kundu$^{1,2}$}

\date{}
\maketitle
%\enlargethispage{1 in.}
\begin{abstract}
The progressive censoring scheme has received considerable amount of attention in the last fifteen years.  During the last few 
years joint progressive censoring scheme has gained some popularity.  Recently, the authors Mondal and Kundu (``A new two sample Type-II 
progressive censoring scheme'', arXiv:1609.05805) introduced a balanced two sample Type-II progressive censoring scheme and provided
the exact inference when the two populations are exponentially distributed.  
In this article we consider the case when the two populations follow Weibull distributions with the common shape parameter and 
different scale parameters.  We obtain the maximum likelihood estimators of the unknown parameters.  It is observed that the 
maximum likelihood estimators cannot be obtained in explicit forms, hence, we propose approximate maximum likelihood estimators, 
which can be obtained in explicit forms.  We construct the asymptotic and bootstrap confidence intervals of the population 
parameters.  Further we derive an exact joint confidence region of the unknown parameters.  We propose an objective function 
based on the expected volume of this confidence set and using that we obtain the optimum progressive censoring scheme.  Extensive
simulations have been performed to see the performances of the proposed method, and one real data set has been analyzed 
for illustrative purposes.     
\end{abstract}



\noindent $^1$ Department of Mathematics and Statistics, Indian Institute of
Technology Kanpur, Pin 208016, India.  

\noindent ${1,2}$ Corresponding author.  E-mail: kundu@iitk.ac.in, Phone no. 91-512-2597141, Fax no. 91-512-2597500.

\newpage

\section{\sc Introduction}

In any life testing experiment censoring in inevitable.  Different censoring schemes have been introduced in the literature 
to optimize time, cost and efficiency.  Among different censoring schemes, Type-I and Type-II are the two most popular censoring
schemes.  But none of these censoring schemes allows removal of units during life testing experiment.  Progressive censoring 
scheme incorporates this flexibility in a life testing experiment.  Progressive Type-II censoring scheme  allows removal of experimental units during the experiment as well as ensures a certain number of failure to be observed during the experiment 
to make it efficient.  Extensive work had been done on the different aspects of the progressive censoring since the introduction of the
book by Balakrishnan and Aggarwala \cite{balakrishnan2000progressive}.  A comprehensive collection of different work related to
progressive censoring scheme can be found in a recent book by Balakrishnan and Cramer \cite{balakrishnan2014art}.  

But all these development are mainly based on a single population.  Recently two sample joint censoring schemes are becoming 
popular for a life testing experiment mainly to optimize time and cost.  In a Type-II joint censoring scheme two samples are 
put on a life testing experiment simultaneously and the experiment is continued until a certain number of failures are observed. 
Balakrishnan and Rasouli \cite {balakrishnan2008exact} first considered the likelihood inference for two exponential 
populations under joint a Type-II censoring scheme.  Ashour and Eraki \cite{ashour2014parameter} extended the results for multiple 
populations and when the lifetime of different populations follow Weibull distributions.

Recently, Rasouli and Balakrishnan \cite{rasouli2010exact} introduced a joint progressive Type-II censoring (JPC) scheme and 
provided the exact likelihood inference for two exponential populations under this censoring scheme.  Parsi and Ganjali 
\cite{parsi2011conditiona} extended the results of Rasouli and Balakrishnan \cite{rasouli2010exact} for two Weibull populations.
Doostparast and Ahmadi et al. \cite{doostparast2013bayes} provided the Bayesian inference of the unknown parameters based on the 
data obtained from a JPC scheme under LINEX loss function.    Balakrishnan and Su et al. \cite{balakrishnan2015exact} extended 
the JPC model to general $K$ populations and studied exact likelihood inference of the unknown parameters for exponential 
distributions.

Mondal and Kundu \cite{MK2016} recently introduced a balanced joint progressive Type-II censoring (BJPC) scheme  and it is 
observed 
that it has certain advantages over the JPC scheme originally introduced by Rasouli and Balakrishnan \cite{rasouli2010exact}.
The scheme proposed by Mondal and Kundu \cite{MK2016} has a close connection with the self relocating design proposed by 
Srivastava \cite{srivastava1987}.  
Mondal and Kundu \cite{MK2016} provided the exact likelihood inference for the two exponential populations under a 
BJPC scheme.  The main aim of this paper is to study likelihood inference of two Weibull populations under this new scheme.
We provide the maximum likelihood estimators (MLEs) of the unknown parameters, and it is observed that the MLEs of the unknown
parameters cannot be obtained in explicit form.  Due to this reason we propose to use approximate maximum likelihood 
estimators (AMLEs) of the unknown parameters, which can be obtained in explicit forms.  We propose to use the asymptotic 
distribution of the MLEs and bootstrap method to construct confidence intervals (CI) of the unknown parameters.  We have 
also provided an exact joint confidence region of the parameter set.  Further, we propose an objective function based on the 
expected volume of this confidence set and this has been used to find the optimum censoring scheme (OCS).  Extensive simulations
have been performed to see the effectiveness of the different methods, and one real data set has been analyzed for illustrative 
purposes.

Rest of the paper is organized as follows.  In Section 2 we briefly describe the model and provide necessary assumptions.  
The MLEs and AMLEs are derived in Section 3.  In Section 4 we provide the joint 
confidence region of the unknown parameters.  Next we propose the objective function in Section 5.  In Section 6 we provide the
simulation results and the analysis of a real data set.  Finally we conclude the paper in Section 7.  
  
\section{\sc Model Description and Model Assumption}

The balanced joint Type-II progressive censoring scheme proposed by Mondal and Kundu \cite{MK2016} can be briefly described as follows.
Suppose there are two lines of similar products and it is important to study the relative merits of these two products.
A sample of size $m$ is drawn from one product line (say A) and another sample of size $m$ is drawn from the other product 
line (say B).  Let $k$ be the total number of failures to be observed from the life testing experiment and $R_1$,\ldots,$R_{k-1}$  are 
pre-specified non-negative integers satisfying $\sum_{i=1}^{k-1}(R_i+1)< m$.   Under the BJPC scheme,  two sets of samples from 
these two products are simultaneously put on a test.  Suppose the first failure is coming from the product line A and the first 
failure time is denoted by $W_1$, then at $W_1$, $R_1$ units are removed randomly from the remaining $m-1$ surviving units 
of the product line A as well as $R_1+1$ units are chosen randomly from the remaining $m$ surviving units of product line B 
and they are removed from the experiment.  Next, if the second failure is coming from the product line B at time point $W_2$, $R_2+1$ units are withdrawn from the remaining $m-R_1-1$ units from the product line A and $R_2$ units withdrawn from the 
remaining $m-R_1-2$ units from the product line B randomly at $W_2$.  The test is continued until $k$ failures are observed with removal of all the remaining surviving units from both the product lines at the $k-$th failure.  In this life testing experiment a new set of random variable $Z_1$,\ldots,$Z_k$ is introduced where $Z_i=1$ or $0$ if $i$th failure comes from the product line A  or B respectively.  Under the BJPC, the  data consists of 
$(\bold W, \bold Z)$ where ${\bold W} = (W_1,\ldots W_k)$ and ${\bold Z} = (Z_1,\ldots,Z_k)$.  A schematic diagram of the BJPC is provided in Figure 
\ref{fig-1} and Figure \ref{fig-2}.

\begin{figure}
\begin{tikzpicture}[scale=0.8]

\draw[gray, thick] (-7,4)node[anchor=north]{\textbf{Population 1}} -- (2,4);
%\draw[gray,thick](2,4)--(3,5)--(2.7,3.3)--(3.5,4)--(10,4);
\draw[gray, thick](2,4)--(2.5,3.5)--(2.8,4.5)--(3.1,4)--(10,4);
\draw[gray, thick] (-7,0)node[anchor=north]{\textbf{Population 2}} -- (2,0);
\draw[gray, thick](2,0)--(2.5,-.5)--(2.8,.5)--(3.1,0)--(10,0);
\filldraw[black] (-7,4) circle (2pt) node[anchor= south] {\small start};
\filldraw[black] (-7,0) circle (2pt) node[anchor= south] {\small start};
\draw[dashed,->] (-5,0) --(-4,2) node[anchor=west] {\small $R_1+1$};
\draw[arrows=->] (-5,4)--(-4,6)node[anchor=west] {\small $R_1$};
\filldraw[black](-4,1.5)node[anchor=west]{\small withdrawn};
\filldraw[black](-4,5.5)node[anchor=west]{\small withdrawn};
\draw [dashed] (-5,-1)--(-5,5);
\filldraw[black] (-5,4) circle (2pt) node[anchor= north west] {$W_1$};
\draw[arrows=->] (-1,0) --(0,2) node[anchor=west]{\small $R_2$};
\filldraw[black] (-1,0) circle(2pt) node[anchor=north west]{$W_2$};
\draw[dashed,->] (-1,4)--(0,6) node[anchor=west]{\small $R_2+1$};
\filldraw[black](0,1.5)node[anchor=west]{\small withdrawn};
\filldraw[black](0,5.5)node[anchor=west]{\small withdrawn};
\draw [dashed] (-1,-1)--(-1,5);
\draw[dashed,->] (5,0) --(6,2) node[anchor=west]{\small $m-\sum_{j=1}^{k-1}(R_j+1)$};
\draw[arrows=->] (5,4)--(6,6) node[anchor=west]{\small $m-\sum_{j=1}^{k-1}(R_j+1)-1$};
\draw [dashed] (5,-1)--(5,5);
\filldraw[black] (5,4) circle (2pt) node[anchor= north west] {$W_k$};
\filldraw[black](6,1.5)node[anchor=west]{\small withdrawn};
\filldraw[black](6,5.5) node[anchor=west]{\small withdrawn};

\end{tikzpicture}
\caption{ $k$th failure comes from Population 1}   \label{fig-1}
\begin{tikzpicture}[scale=0.8]

\draw[gray, thick] (-7,4)node[anchor=north]{\textbf{Population 1}} -- (2,4);
%\draw[gray,thick](2,4)--(3,5)--(2.7,3.3)--(3.5,4)--(10,4);
\draw[gray, thick](2,4)--(2.5,3.5)--(2.8,4.5)--(3.1,4)--(10,4);

\draw[gray, thick] (-7,0)node[anchor=north]{\textbf{Population 2}} -- (2,0);
\draw[gray, thick](2,0)--(2.5,-.5)--(2.8,.5)--(3.1,0)--(10,0);

\filldraw[black] (-7,4) circle (2pt) node[anchor= south] {\small start};
\filldraw[black] (-7,0) circle (2pt) node[anchor= south] {\small start};

\draw[dashed,->] (-5,0) --(-4,2) node[anchor=west] {\small $R_1+1$ };
\draw[arrows=->] (-5,4)--(-4,6)node[anchor=west] {\small $R_1$};
\draw [dashed] (-5,-1)--(-5,5);
\filldraw[black] (-5,4) circle (2pt) node[anchor= north west] {$W_1$};
\filldraw[black](-4,1.5)node[anchor=west]{\small withdrawn};
\filldraw[black](-4,5.5)node[anchor=west]{\small withdrawn};
\draw[arrows=->] (-1,0) --(0,2) node[anchor=west]{\small $R_2$};
\filldraw[black] (-1,0) circle(2pt) node[anchor=north west]{$W_2$};
\draw[dashed,->] (-1,4)--(0,6) node[anchor=west]{\small $R_2+1$};
\draw [dashed] (-1,-1)--(-1,5);
\filldraw[black](0,1.5)node[anchor=west]{\small withdrawn};
\filldraw[black](0,5.5)node[anchor=west]{\small withdrawn};

\draw[arrows=->] (5,0) --(6,2) node[anchor=west]{\small $m-\sum_{j=1}^{k-1}(R_j+1)-1$};
\filldraw[black] (5,0) circle (2pt) node[anchor= north west] {$W_k$};

\draw[dashed,->] (5,4)--(6,6) node[anchor=west]{\small $m-\sum_{j=1}^{k-1}(R_j+1)$};
\draw [dashed] (5,-1)--(5,5);
\filldraw[black](6,1.5)node[anchor=west]{\small withdrawn};
\filldraw[black](6,5.5)node[anchor=west]{\small withdrawn};
\end{tikzpicture}
\caption{$k$th failure comes from Population 2}   \label{fig-2}
\end{figure}

A random variable $X$ is said to follow Weibull distribution with the shape parameter $\alpha > 0$ and the scale parameter
$\lambda > 0$ if it has the following probability density function (PDF)
\be
f(x; \alpha, \lambda) = \left \{ \begin{array}{lll}
\alpha \lambda x^{\alpha-1} e^{-\lambda x^{\alpha}} & \hbox{if} & x > 0  \cr
0 & \hbox{if} & x \le 0,  
\end{array}
\right .
\ee
and it will be denoted by WE($\alpha, \lambda$).  
We assume the lifetimes of $m$ units of product line A, say $X_1,\ldots,X_m$, are independent identically distributed (i.i.d) 
random variables from WE($\alpha,\lambda_1$) and the lifetimes of $m$ units of product line B, say $Y_1,\ldots ,Y_m$ are i.i.d 
random variables from WE($\alpha,\lambda_2$).  

\section{\sc Point Estimations}

\subsection{\sc Maximum Likelihood Estimators (MLEs)}

The likelihood function of the unknown parameters $(\alpha, \lambda_1, \lambda_2)$ based on the observed data $(\bold W,\bold Z)$,
is given by
\bea
L(\alpha,\lambda_1,\lambda_2|\bold w,\bold z) & = &  C{\alpha}^k {\lambda_1}^{k_1} {\lambda_2}^{k_2} \prod_{i=1}^{k}{w_i}^{\alpha-1} e^{-(\lambda_1+\lambda_2 )A(\alpha)}
\eea
where 
\beanno 
A(\alpha) & = & \sum_{i=1}^{k}c_iw^{\alpha}_i, \ \ \ c_i=R_i+1; i = 1, \ldots, k-1,  \\
c_k & = & m-\sum_{i=1}^{k-1}(R_i+1), \ \ \ k_1=\sum_{i=1}^{k}z_i, \ \ \ k_2=\sum_{i=1}^{k}(1-z_i)=k-k_1;  \\
C & = & \prod_{i=1}^{k}(m-\sum_{j=1}^{i-1}(R_j+1)).
\eeanno
The log-likelihood function without the normalizing constant is given by
\begin{equation}
l(\alpha,\lambda_1,\lambda_2|\bold w,\bold z) =k\ln(\alpha) + k_1\ln(\lambda_1)+k_2\ln(\lambda_2)- (\lambda_1+\lambda_2)A(\alpha)
+ (\alpha-1)\sum_{i=1}^{k} \ln(w_i).    \label{ll-func}
\end{equation}
Hence, the normal equations can be obtained by taking partial derivatives of the log-likelihood function (\ref{ll-func}) and equating 
them to zero as given below
\bea
\frac{\partial l}{\partial \lambda_1} & = & \frac{k_1}{\lambda_1}-\sum_{i=1}^{k}c_i{w_i}^{\alpha}=0,   \label{eq1}  \\
\frac{\partial l}{\partial \lambda_2} & = & \frac{k_2}{\lambda_2}-\sum_{i=1}^{k}c_i{w_i}^{\alpha}=0,   \label{eq2}  \\
\frac{\partial l}{\partial \alpha} & = & \frac{k}{\alpha} - (\lambda_1+\lambda_2)\sum_{i=1}^{k}c_i\ln(w_i){w_i}^{\alpha} + \sum_{i=1}^{k}\ln{w_i} = 0.
\label{eq3}
\eea
For a given $\alpha$, when $k_1>0$ and $k_2>0$ the MLEs of $\lambda_1$ and $\lambda_2$ can be obtained from (\ref{eq1}) and (\ref{eq2}) as
follows:
$$
\widehat{\lambda}_1(\alpha) = \frac{k_1}{A(\alpha)} \ \ \ \hbox{and} \ \ \ \ \widehat{\lambda}_2(\alpha)=\frac{k_2}{A(\alpha)}.
$$
When $\alpha$ is also unknown, it is possible to obtain the MLE of $\alpha$ from (\ref{eq3}) by substituting $\lambda_1$ and $\lambda_2$ 
with $\ds \widehat{\lambda}_1(\alpha)$ and  $\ds \widehat{\lambda}_2(\alpha)$, respectively.  Alternatively, 
the MLE of $\alpha$ can be obtained by  maximizing the profile log-likelihood function of $\alpha$, 
$\ds l(\alpha, \widehat{\lambda}_1(\alpha), \widehat{\lambda}_2(\alpha)) = P(\alpha)$ (say), 
where 
\be
P(\alpha)=k\ln(\alpha) -k\ln A(\alpha) +(\alpha-1)\sum_{i=1}^{k}\ln(w_i).   \label{palpha}
\ee
We need the following result for further development.

\noindent {\sc Lemma 1:} The function $P(\alpha)$ as defined in (\ref{palpha}) attains a unique maximum at some 
$\alpha^*$  $\in$ $(0,\infty)$ where $\alpha^*$ is the unique solution of
\begin{equation}
\frac{1}{\alpha} - H(\alpha) +\frac{1}{k}\sum_{i=1}^{k}\ln(w_i)=0,   \label{ne}
\end{equation}
where \hspace{.3mm}$\ds H(\alpha)=\frac{A'(\alpha)}{A(\alpha)}=\frac{\sum_{i=1}^{k}c_i\ln(w_i)w^{\alpha}_i}
{\sum_{i=1}^{k}c_i w^{\alpha}_i}$.

\noindent {\sc Proof:} See in the Appendix.   \qed

Once the unique MLE of $\alpha$, say $\widehat{\alpha}_{MLE}$, is obtained as a solution of (\ref{ne}), then the MLEs of
$\lambda_1$ and $\lambda_2$ also can be obtained uniquely as $\ds \widehat{\lambda}_{1_{MLE}} = \widehat{\lambda}_1(\widehat{\alpha}_{MLE})$
and $\ds \widehat{\lambda}_{2_{MLE}} = \widehat{\lambda}_2(\widehat{\alpha}_{MLE})$, respectively, provided $k_1 > 0$ and $k_2 > 0$.

\subsection{\sc Approximate Maximum Likelihood Estimators}

Since the MLEs cannot be obtained in explicit forms, we propose to use approximate MLEs (AMLEs) of the unknown parameters 
which can be obtained in explicit forms.  They are obtained by expanding the normal equations using Taylor series expansion of 
first order.  It can be easily seen that for $i = 1, 2, \ldots, k$, the distribution of $(\lambda_1+\lambda_2){W_i}^{\alpha}$ is independent of the parameters $\alpha$, $\lambda_1$, 
$\lambda_2$ (see Lemma 2 in Section 4).  Let us define the following random variables: 
$$
U_i = \ln \ ((\lambda_1+\lambda_2){W_i}^{\alpha}) \ \ \ \hbox{and} \ \ \ \ V_i = \ln W_i; \ \ \ i = 1, 2, \ldots, k.  
$$
Therefore, if $\theta = \ln \ (\lambda_1+\lambda_2)$, then 
$\ds U_i = \alpha V_i + \theta$,  and the distribution of $U_i$ is free from $\alpha$, $\lambda_1$, $\lambda_2$, 
for $i = 1, 2, \ldots k$.  

Now from (\ref{eq1}) and (\ref{eq2}) using $u_i$ and $v_i$ as defined above,  we obtain 
\be \label{eq4}
\sum_{i=1}^{k}c_ie^{u_i}=k,
\ee
and from (\ref{eq3})  we have 
\be \label{eq5}
\sum_{i=1}^{k}c_i v_i e^{u_i} = \frac{k}{\alpha} + \sum_{i=1}^{k}v_i. 
\ee 
Using Taylor series expansion of order 1 of $e^{u_i}$, we obtain the AMLEs of the unknown parameters as follows.  The AMLE of $\alpha$, 
say ${\widehat{\alpha}}_{AMLE}$, is the positive root of
\begin{equation}
{\alpha}^2 \left (E_1 - \frac{D_1E_2}{D_2} \right ) + \alpha \left (E_3+ \frac{D_3E_2}{D_2} \right ) = k,
\end{equation}
and the AMLEs of $\lambda_1$, $\lambda_2$ are given by 
$$
\widehat{\lambda}_{1,AMLE} = \frac{k_1}{A({\widehat{\alpha}}_{AMLE})}, \ \ \ 
\widehat{\lambda}_{2,AMLE} = \frac{k_2}{A({\hat{\alpha}}_{AMLE})},
$$
respectively.  Here,
$$
E_1 = \sum_{i=1}^{k}c_iA_i{v_i}^2, \ \ \ E_2=\sum_{i=1}^{k}c_iA_iv_i, \ \ \ \ E_3=\sum_{i=1}^{k}(c_iB_i-1)v_i,
$$
$$
D_1 = \sum_{i=1}^{k}c_iA_iv_i, \ \ \ D_2 = \sum_{i=1}^{k}c_iA_i, \ \ \ \ D_3=k-\sum_{i=1}^{k}c_iB_i,
$$
and for $\ds \xi_i = E(U_i)$, 
$$
A_i = e^{\xi_i} \ \ \ \ B_i=e^{\xi_i}(1-\xi_i); \ \ \ i = 1, 2, \ldots, k.
$$

\section{\sc Exact Confidence Set}

In this section we provide a methodology to construct an exact 100$(1-\gamma)$\% confidence set of $\alpha$, $\lambda_1$ and 
$\lambda_2$.  We need the following results for further development.

\noindent {\sc Lemma 2:} Suppose $G_1, G_2, \ldots, G_k$ are independent exponential random variables and 
$$
E(G_j) = \frac{1}{(\lambda_1 + \lambda_2)(m-\sum_{l=1}^{j-1}(R_l+1))}; \ \ \ j = 1, 2, \ldots, k.
$$
Then $\ds W_i^{\alpha} \overset{d}{=}\sum_{j=1}^{i}G_j$ for $i = 1, 2, \ldots, k$, where $W_i$'s are same as defined before, and
$`\overset{d}{=}$' means equal in distribution.

\noindent {\sc Proof:} See in the Appendix.   \qed

Let us use the following transformation:
\beanno
S_1 & = & m(\lambda_1+\lambda_2){W_1}^{\alpha}\\
S_2 & = & (m-(R_1+1))(\lambda_1+\lambda_2)({W_2}^{\alpha}-{W_1}^{\alpha})\\
\vdots &  & \\
S_k & = & (m-\sum_{i=1}^{k-1}(R_i+1))(\lambda_1+\lambda_2)({W_k}^{\alpha}-{W_{k-1}}^{\alpha}).
\eeanno
From Lemma 2, it is evident that $S_1, S_2, \ldots,S_k$ are independent identically distributed (i.i.d.) exponential random 
variables with mean one.   Let us define 
$$
U = 2 \sum_{i=2}^{k}S_i, \ \ \ V=2S_1, \ \ \ T_1=\frac{U}{(k-1)V}, \ \ \ T_2=U+V.
$$
Observe that $U$ and $V$ are independent, $U \sim {\chi}_{2k-2}^2$, $V\sim {\chi}_{2}^2$, $T_1 \sim F_{2k-2,2}$  and  
$T_2 \sim \chi^2_{(2k)}$.  Using Basu's theorem it follows that $T_1$ and $T_2$ are independently distributed.   Note that 
\bea
T_1 & = & \frac{U}{(k-1)V} = \frac{\sum_{i=1}^k S_i}{S_1(k-1)} - \frac{1}{k-1} = \frac{\sum_{i=1}^k c_i W_i^{\alpha}}{(k-1)m W_1^{\alpha}}
- \frac{1}{k-1}.   \label{mod-t}  \\
T_2 & = & = 2(\lambda_1+\lambda_2) \sum_{i=1}^{k} c_iW^{\alpha}_i.  \label{mod-t2}
\eea
From (\ref{mod-t}) it is clear that $T_1$ is a function $\alpha$, and from now on we denote it by $T_1(\alpha)$.  We have the 
following result.
 
\noindent {\sc Lemma 3:} Let $0 < w_1 < w_2 < \ldots < w_k$, and 
$$
t_1(\alpha) = \frac{\sum_{i=1}^k c_i w_i^{\alpha}}{(k-1)m w_1^{\alpha}} - \frac{1}{k-1},
$$
then $t_1(\alpha)$ is a strictly increasing function in  $\alpha$ and $\lim_{\alpha \rightarrow 0} t_1(\alpha)$ = 0, 
$\lim_{\alpha \rightarrow \infty} t_1(\alpha)$ = $\infty$.  Hence, the equation $t_1(\alpha)=t$ has a unique solution for 
$\alpha>0$ and for all $t > 0$.

\noindent {\sc Proof:} See Lemma 1 in Wu and Shuo-Jye \cite{wu2002estimations}.   \qed

We introduce the following notations.  Let $\varphi(\cdot) = t_1^{-1}(\cdot)$ and note that $\varphi(\cdot)$ is an increasing 
function.  $F_{\gamma,\delta_1,\delta_2}$ denotes the upper $\gamma$-th quantile of $F$ distribution with degrees of freedom 
$\delta_1$, $\delta_2$ and $\chi^2_{\gamma, \delta}$ denotes the upper $\gamma$-th quantile of $\chi^2$ distribution with degrees of freedom $\delta$. 
\begin{theorem}  \label{thm-4p1}

\begin{itemize}
\item[(i)] A $100(1-\gamma)\%$ confidence interval of
$\alpha$ is given by 
$$
\left [ \varphi(F_{1-\gamma/2, 2k-2,2}), \ \ \varphi(F_{\gamma/2, 2k-2,2}) \right ] = B({\gamma}) \ \ \ \ (\hbox{say}).
$$ 
\item[(ii)] For a given $\alpha$, a $100(1-\gamma)\%$ confidence set of $(\lambda_1, \lambda_2)$ is given by
$$
\left \{(\lambda_1, \lambda_2); \lambda_1 \ge 0, \lambda_2 \ge 0, \frac{\large \chi^2_{1-\gamma/2, 2k}}{2 \sum_{i=1}^k c_i w_i^{\alpha}} 
<\lambda_1+\lambda_2 < \frac{\large \chi^2_{\gamma/2, 2k}}{2 \sum_{i=1}^k c_i w_i^{\alpha}} \right \} = C(\gamma; \alpha) \ \ \ \ (\hbox{say}).
$$

\end{itemize}
\end{theorem}

\noindent {\sc Proof:} 
\begin{itemize}

 \item[({\it i})] Since $T_1(\alpha)\sim F_{2k-2,2}$, we have $P(F_{1-\gamma/2, 2k-2,2} < T_1(\alpha) <F_{\gamma/2, 2k-2,2})=1-\gamma$.  
As $t_1(\alpha)$ is an increasing function of $\alpha$ and $\varphi(t)$ is the unique solution of $t_1(\alpha)=t$ we also have
$$
P(\varphi(F_{1-\gamma/2, 2k-2,2})<\alpha <\varphi(F_{\gamma/2, 2k-2,2}))=1-\gamma.
$$

\item[({\it ii})] Since $T_2 \sim \chi^2_{2k}$, using (\ref{mod-t2}), we obtain 
$$
P( \chi^2_{1-\gamma/2, 2k} < 2 (\lambda_1+\lambda_2) \sum_{i=1}^k c_i W_i^{\alpha} < \chi^2_{\gamma/2, 2k}) = 1-\gamma.
$$
Hence, 
$$
P \left (
(\lambda_1, \lambda_2); \lambda_1 \ge 0, \lambda_2 \ge 0, \frac{\large \chi^2_{1-\gamma/2, 2k}}{2\sum_{i=1}^k c_i w_i^{\alpha}} 
<\lambda_1+\lambda_2 < \frac{\large \chi^2_{\gamma/2, 2k}}{2\sum_{i=1}^k c_i w_i^{\alpha}} \right )=1-\gamma.
$$
\end{itemize} 

\qed

\noindent Note that $C(\gamma; \alpha)$ is a trapezoid enclosed by four straight lines 
$$
i) \ \lambda_1 = 0, \ \ \ \ ii) \ \lambda_2 = 0, \ \ \ \ \ \ iii) \ \lambda_1 + \lambda_2 =  
\frac{\large \chi^2_{1-\gamma/2, 2k}}{2\sum_{i=1}^k c_i w_i^{\alpha}} \ \ \ \ 
iv) \ \lambda_1 + \lambda_2 =  \frac{\large \chi^2_{\gamma/2, 2k}}{2\sum_{i=1}^k c_i w_i^{\alpha}}.
$$

\begin{corollary}  \label{cor-4p1}
A $100(1-\gamma)\%$ joint confidence region of $\alpha$, $\lambda_1$, $\lambda_2$ is given by 
$$
D(\gamma) = \Big\{(\alpha, \lambda_1, \lambda_2); \alpha \in B(\gamma_1), \hspace{2mm} (\lambda_1,\lambda_2) \in 
C(\gamma_2; \alpha) \Big\},
$$
here $\gamma_1$ and $\gamma_2$ are such that $1-\gamma = (1-\gamma_1)(1-\gamma_2)$.
\end{corollary}

\section{\sc Optimum Censoring Scheme }

Finding an optimum censoring scheme is an important problem in any life testing experiment.  In this section we propose a new 
objective function and based on which we provide an algorithm to find the optimum censoring scheme.

In a progressive censoring scheme for fixed sample size ($m$) and for fixed effective sample size ($k$), the {\it efficiency} of the 
estimators
depends on the censoring scheme $\{R_1, \ldots,R_{k-1} \}$.  In practical situation out of all possible set of censoring schemes it is important to find out the optimal censoring scheme (OCS)  i.e. the censoring scheme which provides maximum {\it information} about the unknown parameters.  In this case, for fixed $m$ and $k$, the possible set of censoring schemes consists of $R_i$'s, 
$i=1,\ldots,k-1$ such that $\sum_{i=1}^{k-1}(R_i+1)<m$.
 
In case of Weibull and other lifetime distributions most of the available criteria to find the optimum censoring scheme, are based on 
the expected Fisher information matrix, i.e. the asymptotic variance covariance matrix of the MLEs, see for example 
Ng et al. \cite{ng2004optimal}, Pradhan and Kundu \cite{pradhan2009progressively}, Pradhan and Kundu \cite{pradhan2013inference}, 
Balakrishnan and Cramer \cite{balakrishnan2014art} and the references cited therein.  In this paper we propose a new
objective function based on the volume of the exact confidence set of the unknown parameters, which is more reasonable than the asymptotic
variance covariance matrix.

In this case first we will show that it is possible to determine the volume of the exact joint confidence set of $\alpha$, $\lambda_1$ 
and $\lambda_2$.  From the Theorem \ref{thm-4p1} the area  $Area(C(\gamma_2; \alpha)$ of the trapezoid $C(\gamma_2; \alpha)$ is
$$
Area(C(\gamma_2; \alpha) = \frac{(\chi^2_{\gamma_2/2, 2k})^2 -(\chi^2_{1-\gamma_2/2, 2k})^2}{8 A(\alpha)^2}.
$$
The volume  $V(D(\gamma))$ of the confidence region $D(\gamma)$ as in Corollary \ref{cor-4p1}, becomes
\be
V(D(\gamma)) = \frac{1}{8} ((\chi^2_{\gamma_2/2, 2k})^2 -(\chi^2_{1-\gamma_2/2, 2k})^2) \int_{B(\gamma_1)} \frac{1}{A(\alpha)^2} \,d\alpha.
\label{vol-cr}
\ee
Based on (\ref{vol-cr}), we propose the objective function as  $E_{data}(V(D(\gamma)))$. Therefore, for fixed $m$ and $k$ if 
${\cal R}_1 = (R_{1,1},R_{2,1},\ldots,R_{k-1,1})$ and ${\cal R}_2 =(R_{1,2},R_{2,2},\ldots,R_{k-1,2})$ are two censoring plans then 
${\cal R}_1$ is better than ${\cal R}_2$ if ${\cal R}_1$ provides smaller $E_{data}(V(D(\gamma)))$ than ${\cal R}_2$.
The following algorithm can be used to compute $E_{data}(V(D(\gamma)))$, for fixed $m$, $k$ and $R_1,\ldots,R_{k-1}$.

\noindent {\sc Algorithm:} 

\begin{itemize}
\item{\sc Step1:} Given $m$, $k$ and $R_1,\dots,R_{k-1}$  generate the data $(\bold W, \bold Z)$ under BJPC from two Weibull populations, $WE(\alpha, \lambda_1)$ and $WE(\alpha, \lambda_2)$ .
\item{\sc Step2:} Compute $V(D(\gamma))$ based on the data, this can be done by various numerical method like trapezoidal rule.
\item{\sc Step3:} Repeat Step 1 to 2 say $B$ times, and take their average which approximates $E_{data}(V(D(\gamma)))$.
\end{itemize}

\section{\sc Simulation Study And Data Analysis}

\subsection{\sc Simulation Study}

In this section we compare the performance of the MLEs and AMLEs based on an extensive simulation experiment.  We have taken 
the sample size $m=25$ and different effective sample size namely, $k=15, 20$.  For different censoring schemes and for different 
parameter values we compute the average estimate (AE) and mean square error (MSE) of the MLEs and AMLEs based on 10,000 replications.
We have also computed the 90\% asymptotic and percentile bootstrap confidence intervals, and we have 
reported the average length (AL) and the coverage percentages (CP) in each case.  Bootstrap confidence intervals are obtained based
on 1000 bootstrap samples.  The results are reported in Tables \ref{table-1} to \ref{table-6}.  We have used the following
notations denoting different censoring scheme.  For example, 
the progressive censoring scheme $R_1=2, R_2=0, R_3=0, R_4=0$, has been denoted by $R=(2,0_{(3)})$.

Some of the points are quite clear from this simulation experiment.  It is observed that for fixed $m$ (sample size) as $k$ 
(effective sample size) increases the biases and MSEs decrease for both MLEs and AMLEs as expected.  The performances of the MLEs and
AMLEs are very close to each other in all cases considered both in terms of biases and MSEs.  Hence we recommend to use AMLEs 
in this case as they have explicit forms.  Now comparing the bootstrap and asymptotic confidence intervals in terms of the 
average lengths and coverage percentages, it is observed the performance of the bootstrap confidence intervals are not very 
satisfactory.  Most of the times it cannot maintain the nominal level of the coverage percentages.  Where as even for small 
sample sizes the performances of the asymptotic confidence intervals are quite satisfactory.  In most of the cases considered the 
coverage percentages are very close to the nominal level.  Hence, we recommend to use asymptotic confidence interval in this case.

We have further studied the relation between $E(V(D(\gamma)))$  and the expected time of test (ETOT) for different censoring 
schemes and for different parameter values.  $E(V(D(\gamma))$ is computed as described in Section 5 based on $B=50,000$ 
samples with $\gamma = 0.1$ and $\gamma_1 = \gamma_2$.  The ETOT i.e. $E(W_k)$ is computed by Monte-Carlo simulation based on 10,000 samples.  The results are reported in 
Tables \ref{table-7} to \ref{table-9} for different parameter values and for different censoring schemes.  We have also provided
a scatter plot of ETOT vs. $E(V(D(\gamma)))$ for different censoring schemes in Figure \ref{sfig1}.  
It is evident that as the ETOT increases, $E(V(D(\gamma)))$ decreases as expected.

\begin{table}[H]
\caption{ $m=25,n=25,\alpha=0.5,\lambda_1=0.5, \lambda_2=1$}
\label{table-1}
\begin{center}
%\centering
\scalebox{.7}{
\begin{tabular}{llllll}%\label{table-5}
\toprule
\multicolumn{1}{c}{Censoring scheme} & \multicolumn{1}{c}{Parameter}
&\multicolumn{2}{c}{MLE} & \multicolumn{2}{c}{AMLE}\\
&&\multicolumn{1}{c}{AE} & \multicolumn{1}{c}{MSE}
&\multicolumn{1}{c}{AE}  &\multicolumn{1}{l}{MSE} \\
\midrule
k=15,R=(7,0$_{(13)}$) & $\alpha$ &0.550&0.017&0.534&0.015\\
 & $\lambda_1$ &0.576&0.109&0.568&0.101\\
 & $\lambda_2$ &1.149&0.279&1.132&0.253 \\
 \midrule
 k=15,R=(0$_{(6)}$,7,0$_{(7)}$) &$\alpha$ &0.552&0.019&0.547&0.018\\
 &$\lambda_1$ &0.601&0.143&0.595&0.137\\
 & $\lambda_2$ &1.198&0.371&1.187&0.352\\
 \midrule
 k=15,R=(0$_{(13)}$,7) & $\alpha$ &0.564&0.024&0.559&0.023\\
 & $\lambda_1$ &0.628&0.184&0.622&0.176\\
 & $\lambda_2$ &1.248&0.514&1.236&0.491\\
 \midrule
 k=20,R=(3,0$_{(18)}$) &$\alpha$ &0.537&0.012&0.529&0.011\\
   & $\lambda_1$ &0.547&0.056&0.544&0.054\\
   & $\lambda_2$ &1.079&0.123&1.074&0.118\\
   \midrule
 k=20,R=(0$_{(9)}$,3,0$_{(9)}$) & $\alpha$ &0.539&0.013&0.534&0.012\\
 & $\lambda_1$ &0.548&0.062&0.546&0.061\\
 & $\lambda_2$ &1.097&0.147&1.093&0.143\\
 \midrule
 k=20,R=(0$_{(18)}$,3)& $\alpha$ &0.538&0.012&0.529&0.011\\
 & $\lambda_1$ &0.542&0.055&0.539&0.054\\
 & $\lambda_2$ &1.083&0.130&1.078&0.125\\
 \bottomrule
\end{tabular} 
}
\end{center}
\end{table}

\begin{table}[H]
\caption{ $m=25,n=25,\alpha=1,\lambda_1=0.5, \lambda_2=1$}
\label{table-2}
\begin{center}
%\centering
\scalebox{.7}{
\begin{tabular}{llllll}%\label{table-5}
\toprule
\multicolumn{1}{c}{Censoring scheme} & \multicolumn{1}{c}{Parameter}
&\multicolumn{2}{c}{MLE} & \multicolumn{2}{c}{AMLE}\\
&&\multicolumn{1}{c}{AE} & \multicolumn{1}{c}{MSE}
&\multicolumn{1}{c}{AE}  &\multicolumn{1}{l}{MSE} \\
\midrule
k=15,R=(7,0$_{(13)}$) & $\alpha$&1.096&.071&1.064&0.063 \\
 & $\lambda_1$ &0.575&0.103&0.566&0.095\\
 & $\lambda_2$ &1.154&0.292&1.136&0.264\\
 \midrule
 k=15,R=(0$_{(6)}$,7,0$_{(7)}$) &$\alpha$&1.107&0.078&1.096&0.074\\
 &$\lambda_1$&0.602&0.155&0.597&0.148 \\
 & $\lambda_2$ &1.204&0.446&1.193&0.426\\
 \midrule
 k=15,R=(0$_{(13)}$,7) & $\alpha$ &1.126&0.101&1.116&0.097\\
 & $\lambda_1$ &0.620&0.210&0.614&0.201\\
 & $\lambda_2$ &1.244&0.660&1.232&0.625\\
 \midrule
 k=20,R=(3,0$_{(18)}$) &$\alpha$ &1.082&0.057&1.073&0.055\\
   & $\lambda_1$ &0.557&0.066&0.554&0.064\\
   & $\lambda_2$ &1.109&0.162&1.105&0.158\\
   \midrule
 k=20,R=(0$_{(9)}$,3,0$_{(9)}$) & $\alpha$ &1.080&0.052&1.071&0.050\\
 & $\lambda_1$ &0.550&0.060&0.548&0.059\\
 & $\lambda_2$ &1.093&0.139&1.089&0.135\\
 \midrule
 k=20,R=(0$_{(18)}$,3)& $\alpha$ &1.085&0.058&1.076&0.056\\
 & $\lambda_1$ &0.555&0.066&0.553&0.065\\
 & $\lambda_2$ &1.113&0.172&1.109&0.167\\
 \bottomrule
\end{tabular} 
}
\end{center}
\end{table}

\begin{table}[H]
\caption{ $m=25,n=25,\alpha=2,\lambda_1=0.5, \lambda_2=1$}
\label{table-3}
\begin{center}
%\centering
\scalebox{.7}{
\begin{tabular}{llllll}%\label{table-5}
\toprule
\multicolumn{1}{c}{Censoring scheme} & \multicolumn{1}{c}{Parameter}
&\multicolumn{2}{c}{MLE} & \multicolumn{2}{c}{AMLE}\\
&&\multicolumn{1}{c}{AE} & \multicolumn{1}{c}{MSE}
&\multicolumn{1}{c}{AE}  &\multicolumn{1}{l}{MSE} \\
\midrule
k=15,R=(7,0$_{(13)}$) & $\alpha$ &2.209&0.294&2.147&0.259\\
 & $\lambda_1$ &0.578&0.110&0.569&0.101\\
 & $\lambda_2$ &1.150&0.289&1.132&0.258 \\
 \midrule
 k=15,R=(0$_{(6)}$,7,0$_{(7)}$) &$\alpha$ &2.220&0.319&2.197&0.304\\
 &$\lambda_1$ &0.597&0.132&0.592&0.126\\
 & $\lambda_2$ &1.192&0.388&1.181&0.365\\
 \midrule
 k=15,R=(0$_{(13)}$,7) & $\alpha$ &2.261&0.414&2.240&0.397\\
 & $\lambda_1$ &0.630&0.193&0.624&0.184\\
 & $\lambda_2$ &1.253&0.531&1.241&0.504\\
 \midrule
 k=20,R=(3,0$_{(18)}$) &$\alpha$ &2.148&0.191&2.113&0.178\\
   & $\lambda_1$ &0.545&0.055&0.542&0.054\\
   & $\lambda_2$ &1.087&0.131&1.081&0.125\\
   \midrule
 k=20,R=(0$_{(9)}$,3,0$_{(9)}$) & $\alpha$ &2.158&0.207&2.140&0.199\\
 & $\lambda_1$ &0.548&0.060&0.546&0.059\\
 & $\lambda_2$ &1.098&0.141&1.094&0.137\\
 \midrule
 k=20,R=(0$_{(18)}$,3)& $\alpha$ &2.164&0.227&2.145&0.218\\
 & $\lambda_1$ &0.552&0.063&0.549&0.062\\
 & $\lambda_2$ &1.108&0.155&1.103&0.151\\
 \bottomrule
\end{tabular} 
}
\end{center}
\end{table}

\begin{table}[H]
\caption{ AL and CP of CI's,$m=25,n=25,\alpha=0.5,\lambda_1=0.5, \lambda_2=1$} \label{table-4}
\begin{center}
%\centering
\scalebox{.7}{
\begin{tabular}{llllll}%\label{table-5}
\toprule
\multicolumn{1}{l}{Censoring scheme} & \multicolumn{1}{c}{Parameter} &\multicolumn{2}{l}{Bootstrap 90\% CI} & \multicolumn{2}{l}{Asymptotic 90\%CI}\\
 
&&\multicolumn{1}{c}{AL} & \multicolumn{1}{c}{CP}
&\multicolumn{1}{c}{AL} & \multicolumn{1}{l}{CP} \\
\midrule
k=15,R=(7,0$_{(13)}$) & $\alpha$&0.435&83.1\%&0.378&90.1\% \\
 & $\lambda_1$ &1.141&87.8\%&0.882&90.1\%\\
 & $\lambda_2$ &1.812&84.5\%&1.296&92.1\%\\
 \midrule
 k=15,R=(0$_{(6)}$ ,7,0$_{(7)}$) &$\alpha$&0.457&78.8\%&0.378&89.8\%\\
 &$\lambda_1$&1.374&86.8\%&0.937&90.6\%\\
 & $\lambda_2$ &2.245&83.8\%&1.430&92.9\%\\
 \midrule
 k=15,R=(0$_{(13)}$,7) & $\alpha$ &0.519&79.2\%&0.431&89.7\%\\
 & $\lambda_1$ &1.809&84.1\%&1.049&91.1\%\\
 & $\lambda_2$ &3.084&82.2\%&1.667&93.8\%\\
 \midrule
 k=20,R=(3,0$_{(18)}$) &$\alpha$ &0.365&82.9\%&0.323&89.6\%\\
   & $\lambda_1$ &0.811&88.6\%&0.700&88.3\%\\
   & $\lambda_2$ &1.241&86.4\%&1.018&90.5\%\\
   \midrule
 k=20,R=(0$_{(9)}$,3,0$_{(9)}$) & $\alpha$ &0.366&83.2\%&0.323&89.3\%\\
 & $\lambda_1$ &0.836&89.6\%&0.711&88.8\%\\
 & $\lambda_2$ &1.285&87.5\%&1.044&90.5\%\\
 \midrule
 k=20,R=(0$_{(18)}$,3)& $\alpha$ &0.392&84.7\%&0.343&90.3\%\\
 & $\lambda_1$ &0.852&88.7\%&0.724&89.3\%\\
 & $\lambda_2$ &1.355&85.7\%&1.065&90.8\%\\
 \bottomrule
\end{tabular} 
}
\end{center}
\end{table}

\begin{table}[H]
\caption{ AL and CP of CI's,$m=25,n=25,\alpha=1,\lambda_1=0.5, \lambda_2=1$} \label{table-5}
\begin{center}
%\centering
\scalebox{.7}{
\begin{tabular}{llllll}%\label{table-5}
\toprule
\multicolumn{1}{l}{Censoring scheme} & \multicolumn{1}{c}{Parameter} &\multicolumn{2}{l}{Bootstrap 90\% CI} & \multicolumn{2}{l}{Asymptotic 90\%CI}\\
 
&&\multicolumn{1}{c}{AL} & \multicolumn{1}{c}{CP}
&\multicolumn{1}{c}{AL} & \multicolumn{1}{l}{CP} \\
\midrule
k=15,R=(7,0$_{(13)}$) & $\alpha$&0.866&83.6\%&0.759&89.9\% \\
 & $\lambda_1$ &1.089&89.7\%&0.869&89.4\%\\
 & $\lambda_2$ &1.745&86.2\%&1.293&92.0\%\\
 \midrule
 k=15,R=(0$_{(6)}$ ,7,0$_{(7)}$) &$\alpha$&0.914&78.4\%&0.758&90.0\%\\
 &$\lambda_1$&1.525&86.8\%&0.947&90.5\%\\
 & $\lambda_2$ &2.683&82.8\%&1.451&93.5\%\\
 \midrule
 k=15,R=(0$_{(13)}$,7) & $\alpha$ &1.027&81.7\%&0.862&90.1\%\\
 & $\lambda_1$ &1.639&88.4\%&1.053&91.4\%\\
 & $\lambda_2$ &2.810&84.4\%&1.667&93.7\%\\
 \midrule
 k=20,R=(3,0$_{(18)}$) &$\alpha$ &0.726&82.3\%&0.643&90.3\%\\
   & $\lambda_1$ &0.808&87.2\%&0.697&88.5\%\\
   & $\lambda_2$ &1.222&86.1\%&1.012&90.3\%\\
   \midrule
 k=20,R=(0$_{(9)}$,3,0$_{(9)}$) & $\alpha$ &0.7355&82.5\%&0.648&89.9\%\\
 & $\lambda_1$ &0.835&88.9\%&0.712&89.2\%\\
 & $\lambda_2$ &1.292&86.0\%&1.039&90.7\%\\
 \midrule
 k=20,R=(0$_{(18)}$,3)& $\alpha$ &0.789&81.9\%&0.684&90.2\%\\
 & $\lambda_1$ &0.920&86.7\%&0.723&89.7\%\\
 & $\lambda_2$ &1.419&84.8\%&1.063&90.7\%\\
 \bottomrule
\end{tabular} 
}
\end{center}
\end{table}

\begin{table}[H]
\caption{ AL and CP of CI's,$m=25,n=25,\alpha=2,\lambda_1=0.5, \lambda_2=1$} \label{table-6}
\begin{center}
%\centering
\scalebox{.7}{
\begin{tabular}{llllll}%\label{table-5}
\toprule
\multicolumn{1}{l}{Censoring scheme} & \multicolumn{1}{c}{Parameter} &\multicolumn{2}{l}{Bootstrap 90\% CI} & \multicolumn{2}{l}{Asymptotic 90\%CI}\\
 
&&\multicolumn{1}{c}{AL} & \multicolumn{1}{c}{CP}
&\multicolumn{1}{c}{AL} & \multicolumn{1}{l}{CP} \\
\midrule
k=15,R=(7,0$_{(13)}$) & $\alpha$&1.774&81.4\%&1.515&90.1\% \\
 & $\lambda_1$ &1.169&87.6\%&0.873&89.8\%\\
 & $\lambda_2$ &1.875&85.7\%&1.300&91.9\%\\
 \midrule
 k=15,R=(0$_{(6)}$ ,7,0$_{(7)}$) &$\alpha$&1.813&79.7\%&1.521&89.1\%\\
 &$\lambda_1$&1.332&88.1\%&0.948&90.0\%\\
 & $\lambda_2$ &2.227&83.1\%&1.440&93.0\%\\
 \midrule
 k=15,R=(0$_{(13)}$,7) & $\alpha$ &2.101&78.6\%&1.722&90.0\%\\
 & $\lambda_1$ &1.765&86.4\%&1.074&91.2\%\\
 & $\lambda_2$ &3.010&83.5\%&1.708&93.7\%\\
 \midrule
 k=20,R=(3,0$_{(18)}$) &$\alpha$ &1.461&80.9\%&1.294&89.8\%\\
   & $\lambda_1$ &0.821&88.0\%&0.699&88.9\%\\
   & $\lambda_2$ &1.237&86.6\%&1.014&90.0\%\\
   \midrule
 k=20,R=(0$_{(9)}$,3,0$_{(9)}$) & $\alpha$ &1.478&81.0\%&1.296&89.8\%\\
 & $\lambda_1$ &0.844&88.7\%&0.713&89.0\%\\
 & $\lambda_2$ &1.294&86.2\%&1.044&90.5\%\\
 \midrule
 k=20,R=(0$_{(18)}$,3)& $\alpha$ &1.555&83.3\%&1.374&89.3\%\\
 & $\lambda_1$ &0.883&89.6\%&0.724&89.3\%\\
 & $\lambda_2$ &1.386&86.0\%&1.066&91.5\%\\
 \bottomrule
\end{tabular} 
}
\end{center}
\end{table}

\begin{table}[H]
\caption{ $\alpha=0.5,\lambda_1=0.5, \lambda_2=1$}
\label{table-7}
\begin{center}
%\centering
\scalebox{.7}{
\begin{tabular}{lll}%\label{table-5}
\toprule
\multicolumn{1}{c}{Censoring scheme} & \multicolumn{1}{c}{$E(Vol_{0.1})$} & \multicolumn{1}{c}{$ETOT$} \\
\midrule
m=25,k=20,R=(5,0$_{(18)}$) &12.463&6.420\\
 \midrule
 m=25,k=20,R=(0,5,0$_{(17)}$) &12.583&6.383\\
 \midrule
 m=25,k=20,R=(0$_{(2)}$,5,0$_{(16)}$) &12.845&6.369\\
 \midrule
 m=25,k=20,R=(0$_{(3)}$,5,0$_{(15)}$) &13.032&6.245\\
 \midrule
m=25,k=20,R=(0$_{(4)}$,5,0$_{(14)}$) &13.243&6.181\\
 \midrule
m=25,k=20,R=(0$_{(8)}$,5,0$_{(10)}$) &14.614&6.043 \\
 \midrule
 m=25,k=20,R=(0$_{(14)}$,5,0$_{(4)}$) &17.319&5.092\\
 \midrule
 m=25,k=20,R=(0$_{(16)}$,5,0$_{(2)}$) &20.768&4.458\\
 \midrule
 m=25,k=20,R=(0$_{(17)}$,5,0) &20.918&3.884\\
 \midrule
m=25,k=20,R=($_{(18)}$,5) &22.883&3.023 \\
   \midrule
m=30,k=25,R=(5,0$_{(23)}$) &9.616& 7.181\\ 
 \midrule
 m=30,k=25,R=(0,5,0$_{(22)}$) &9.718&7.145 \\ 
 \midrule
 m=30,k=25,R=(0$_{(2)}$,5,0$_{(21)}$)&9.743&7.081\\
\midrule
m=30,k=25,R=(0$_{(3)}$,5,0$_{(20)}$)&9.834&7.074 \\
\midrule
 m=30,k=25,R=(0$_{(5)}$,5,0$_{(18)}$)&9.908&6.986\\
\midrule
m=30,k=25,R=(0$_{(8)}$,5,0$_{(15)}$)&10.304&6.954\\
\midrule
m=30,k=25,R=(0$_{(12)}$,5,0$_{(11)}$)&10.680&6.675\\
\midrule
m=30,k=25,R=(0$_{(15)}$,5,0$_{(8)}$)&11.217&6.454\\
\midrule
m=30,k=25,R=(0$_{(18)}$,5,0$_{(5)}$)&12.045&5.934\\
\midrule
m=30,k=25,R=(0$_{(23)}$,5)& 14.197 &3.451\\
 \bottomrule
 
\end{tabular} 
}
\end{center}
\end{table}

\begin{table}[H]
\caption{ $\alpha=1,\lambda_1=0.5, \lambda_2=1$}
\label{table-8}
\begin{center}
%\centering
\scalebox{.7}{
\begin{tabular}{lll}%\label{table-5}
\toprule
\multicolumn{1}{c}{Censoring scheme} & \multicolumn{1}{c}{$E(Vol_{0.1})$} & \multicolumn{1}{c}{$ETOT$} \\
\midrule
m=25,k=20,R=(5,0$_{(18)}$) &24.360 &2.393\\
 \midrule
 m=25,k=20,R=(0,5,0$_{(17)}$) &25.214 &2.384\\
 \midrule
 m=25,k=20,R=(0$_{(2)}$,5,0$_{(16)}$) &25.524&2.374\\
 \midrule
 m=25,k=20,R=(0$_{(3)}$,5,0$_{(15)}$) &26.034&2.366\\
 \midrule
m=25,k=20,R=(0$_{(4)}$,5,0$_{(14)}$) &26.513&2.359\\
 \midrule
m=25,k=20,R=(0$_{(8)}$,5,0$_{(10)}$) &28.065&2.300 \\
 \midrule
 m=25,k=20,R=(0$_{(14)}$,5,0$_{(4)}$) &36.513&2.120\\
 \midrule
 m=25,k=20,R=(0$_{(16)}$,5,0$_{(2)}$) &38.831&1.963\\
 \midrule
 m=25,k=20,R=(0$_{(17)}$,5,0) &40.552&1.816\\
 \midrule
m=25,k=20,R=($0_{(18)}$,5) &46.317&1.582 \\
   \midrule
m=30,k=25,R=(5,0$_{(23)}$) &19.331&2.549 \\ 
 \midrule
 m=30,k=25,R=(0,5,0$_{(22)}$) &19.414&2.536 \\ 
 \midrule
 m=30,k=25,R=(0$_{(2)}$,5,0$_{(21)}$)&19.538&2.524\\
\midrule
m=30,k=25,R=(0$_{(3)}$,5,0$_{(20)}$)&19.895&2.523\\
\midrule
 m=30,k=25,R=(0$_{(5)}$,5,0$_{(18)}$)&20.094&2.522\\
\midrule
m=30,k=25,R=(0$_{(8)}$,5,0$_{(15)}$)&20.665&2.488\\
\midrule
m=30,k=25,R=(0$_{(12)}$,5,0$_{(11)}$)&21.478&2.445\\
\midrule
m=30,k=25,R=(0$_{(15)}$,5,0$_{(8)}$)&22.538 &2.374\\
\midrule
m=30,k=25,R=(0$_{(18)}$,5,0$_{(5)}$)&23.846&2.274\\
\midrule
m=30,k=25,R=(0$_{(23)}$,5)& 28.662 &1.692\\
 \bottomrule
 
\end{tabular} 
}
\end{center}
\end{table}

\begin{table}[H]
\caption{ $\alpha=2,\lambda_1=0.5, \lambda_2=1$}
\label{table-9}
\begin{center}
%\centering
\scalebox{.7}{
\begin{tabular}{lll}%\label{table-5}
\toprule
\multicolumn{1}{c}{Censoring scheme} & \multicolumn{1}{c}{$E(Vol_{0.1})$} & \multicolumn{1}{c}{$ETOT$} \\
\midrule
m=25,k=20,R=(5,0$_{(18)}$) &49.927 &1.523\\
 \midrule
 m=25,k=20,R=(0,5,0$_{(17)}$) &50.653 &1.522\\
 \midrule
 m=25,k=20,R=(0$_{(2)}$,5,0$_{(16)}$) &51.265&1.515\\
 \midrule
 m=25,k=20,R=(0$_{(3)}$,5,0$_{(15)}$) &51.578&1.512\\
 \midrule
m=25,k=20,R=(0$_{(4)}$,5,0$_{(14)}$) &52.719&1.508\\
 \midrule
m=25,k=20,R=(0$_{(8)}$,5,0$_{(10)}$) & 57.245&1.492 \\
 \midrule
 m=25,k=20,R=(0$_{(14)}$,5,0$_{(4)}$) &67.359&1.424\\
 \midrule
 m=25,k=20,R=(0$_{(16)}$,5,0$_{(2)}$) &77.601&1.365\\
 \midrule
 m=25,k=20,R=(0$_{(17)}$,5,0) &88.436&1.322\\
 \midrule
m=25,k=20,R=(0$_{(18)}$,5) &89.061& 1.229\\
   \midrule
m=30,k=25,R=(5,0$_{(23)}$) &38.486&1.572 \\ 
 \midrule
 m=30,k=25,R=(0,5,0$_{(22)}$) &38.571&1.572 \\ 
 \midrule
 m=30,k=25,R=(0$_{(2)}$,5,0$_{(21)}$)&38.781 &1.569\\
\midrule
m=30,k=25,R=(0$_{(3)}$,5,0$_{(20)}$)&39.411 &1.568\\
\midrule 
m=30,k=25,R=(0$_{(5)}$,5,0$_{(18)}$)&40.220 &1.561\\
\midrule
m=30,k=25,R=(0$_{(8)}$,5,0$_{(15)}$)&41.074 &1.560\\
\midrule
m=30,k=25,R=(0$_{(12)}$,5,0$_{(11)}$)&43.549 &1.538\\
\midrule
m=30,k=25,R=(0$_{(15)}$,5,0$_{(8)}$)&45.367 &1.517\\
\midrule
m=30,k=25,R=(0$_{(18)}$,5,0$_{(5)}$)&48.084 &1.486\\
\midrule
m=30,k=25,R=(0$_{(23)}$,5)& 55.966 &1.277\\
 \bottomrule
 
\end{tabular} 
}
\end{center}
\end{table}

\begin{figure}[H]
\begin{center}
\includegraphics[scale=0.4]{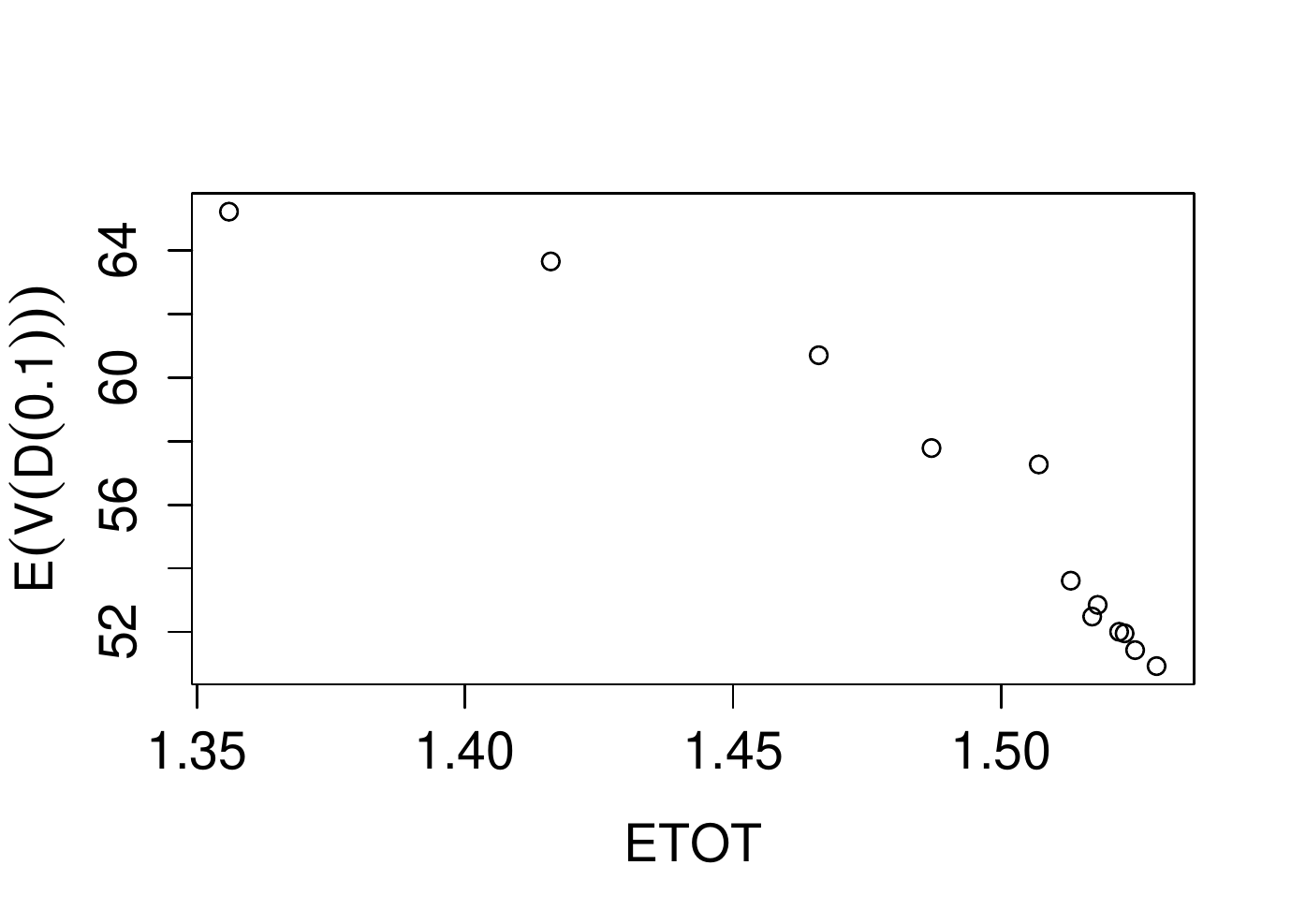}
\caption{The ETOT and $E(D(\gamma))$ for  $m=22,k=20, \alpha=2, \lambda_1=0.5, \lambda_2=1$}
\label{sfig1}
\end{center}
\end{figure}
\subsection{\sc Data Analysis} In this section we perform the analysis of a real data set to illustrate how the propose methods
can be used in practice.  We have used the following data set originally obtained from Proschan  \cite{proschan1963} and here 
the data indicate the failure times (in hour) of air-conditioning system of two airplanes.   The data are provided below.

\noindent Plane 7914: 3, 5, 5, 13, 14, 15, 22, 22, 23, 30, 36, 39, 44, 46, 50, 72, 79, 88, 97, 102, 139,
188, 197, 210.\\
\noindent Plane 7913: 1, 4, 11, 16, 18,18, 24, 31, 39, 46, 51, 54, 63, 68, 77, 80, 82, 97, 106,
141, 163, 191, 206, 216.

From the above data sets we have generated two different jointly progressively censored samples with the censoring schemes
Scheme 1: $k$ = 20 and $R=(14,0_{(8)})$ and Scheme 2: $k=10$, $R=(2_{(7)},0_{(2)})$.  The generated data sets are provided below.

\begin{figure}[htp]
 \centering
\scalebox{.85}{\begin{tabular}{|c c|}
\hline 
     \includegraphics[width=70mm]{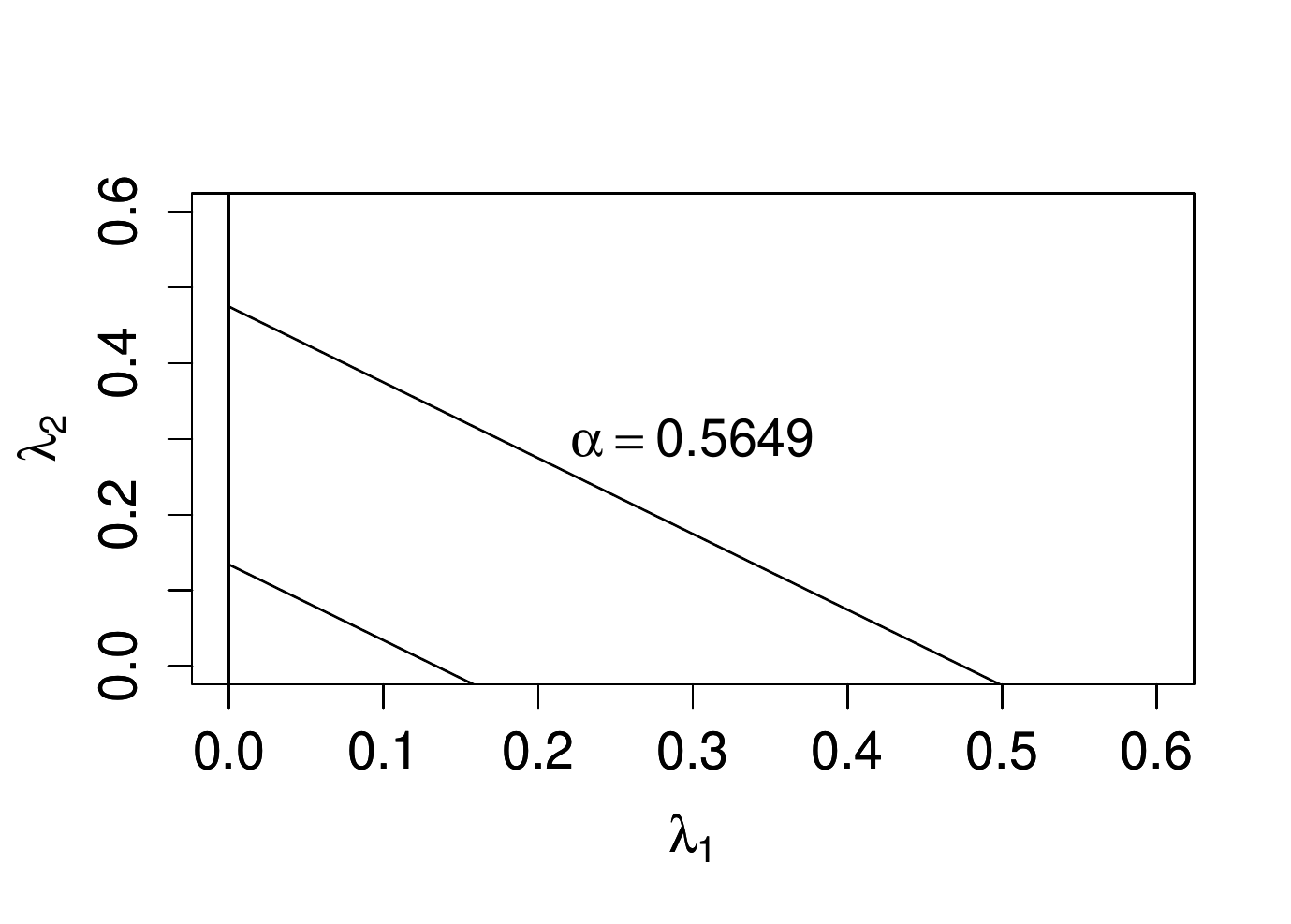}&
     \includegraphics[width=70mm]{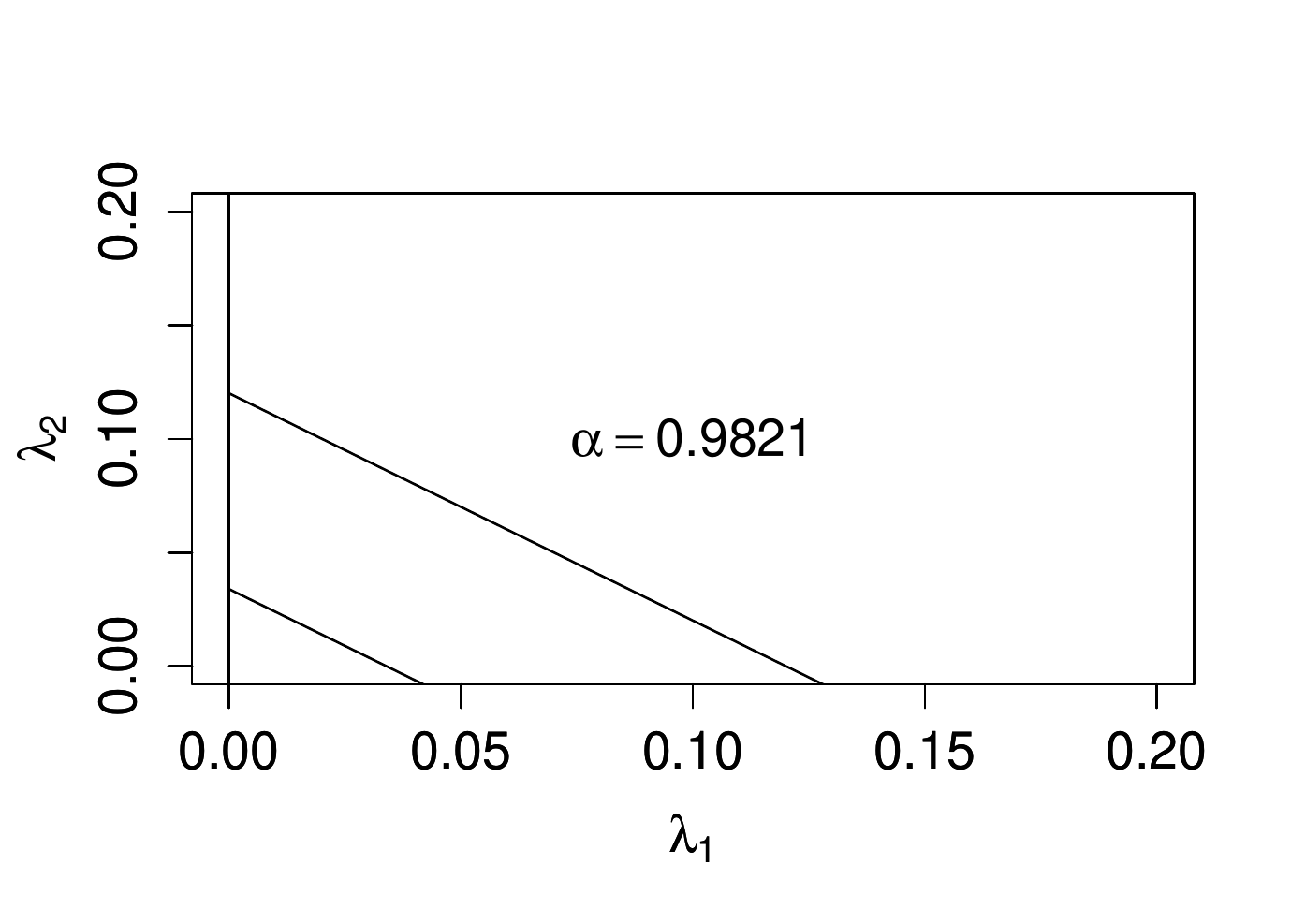}\\
     \includegraphics[width=70mm]{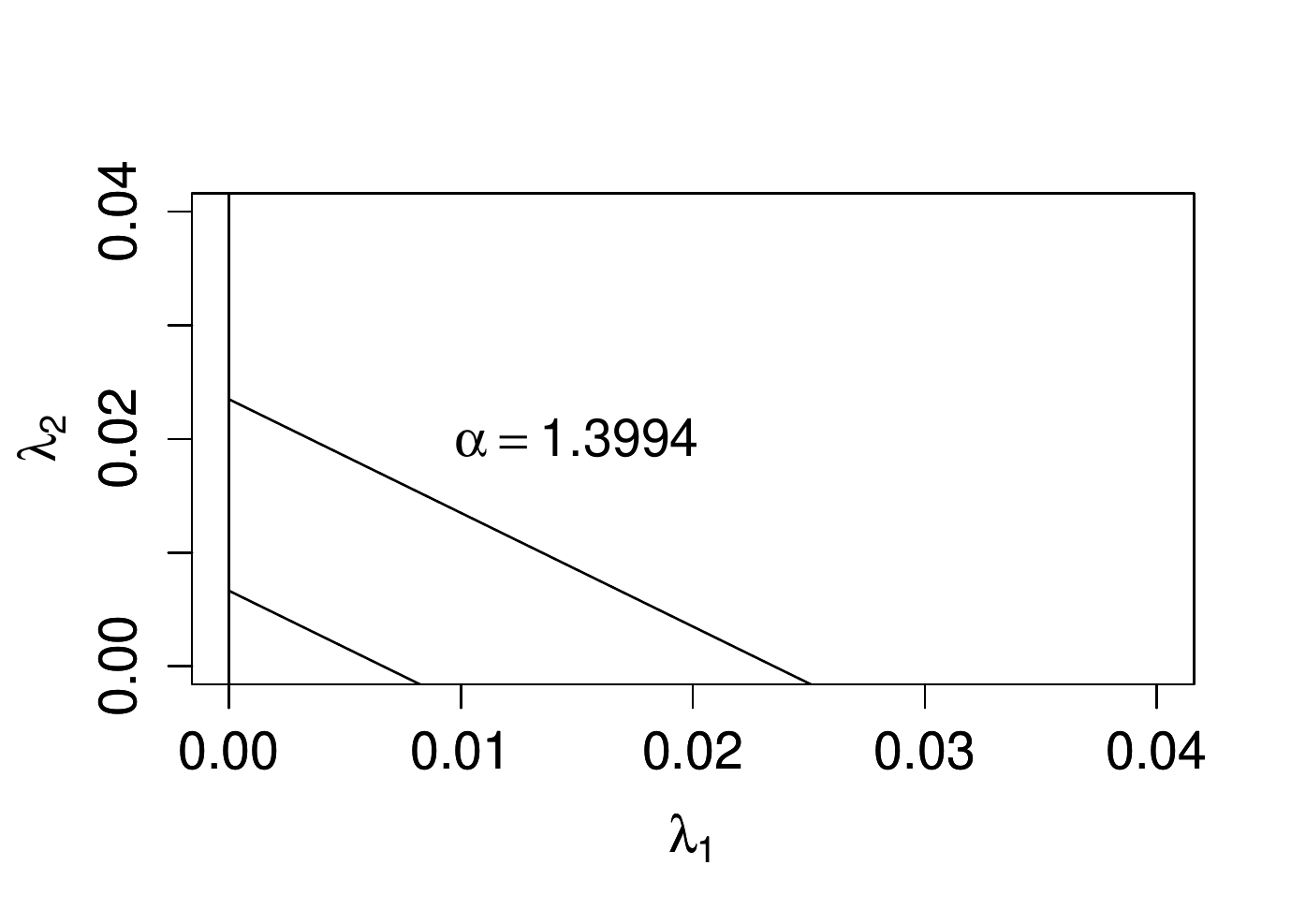}&
     \includegraphics[width=70mm]{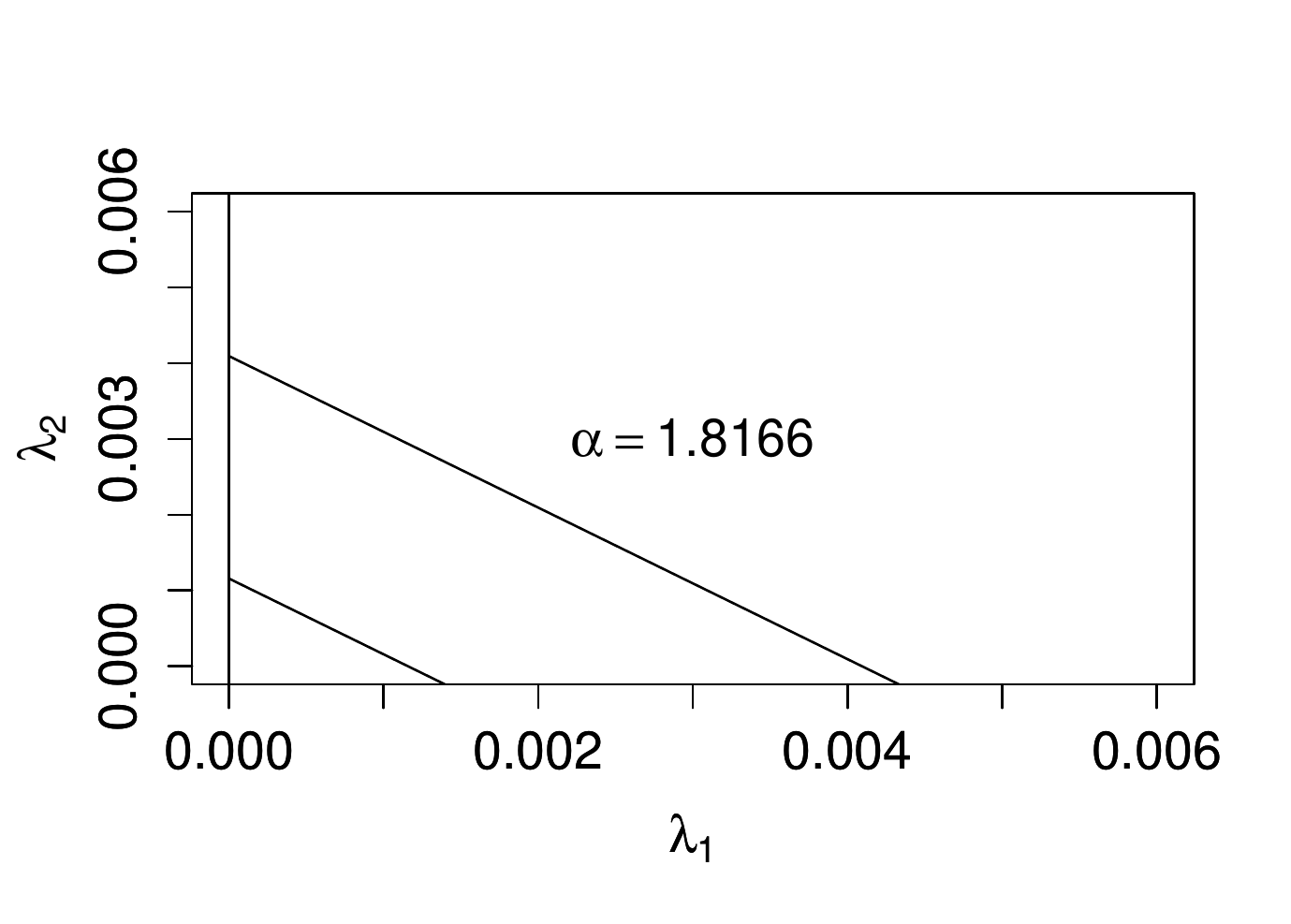}\\
\hline 
\end{tabular}
} 
\caption{Confidence region of $\lambda_1$ and $\lambda_2$ for different values $\alpha$ for scheme-1}
\label{fig-5}
\end{figure}

\noindent {\sc Scheme 1:}
$$
{\bf w}=(1, 4, 5, 13, 15, 16, 22, 36, 80, 97) \ \ \ \ {\bf z}=(0, 0, 1, 1, 1, 0, 1, 1, 0, 0);
$$
For the above data set the MLEs, AMLEs, and the two different 90\% confidence intervals are provided in  Tables \ref{table-10} 
and \ref{table-11}.  In Figure \ref{sfig1} we have provided the profile log-likelihood function $P(\alpha)$ of the 
shape parameter $\alpha$ and it is clear that $P(\alpha)$ attains a unique maximum.  To get an idea about the joint confidence
region of $\alpha$, $\lambda_1$, $\lambda_2$, we have provided the confidence set of $(\lambda_1, \lambda_2)$ for different 
values of $\alpha$ in Figure \ref{fig-5}.

\begin{table}[H]
\caption{ real data analysis(scheme-1)}   
\label{table-10}
\begin{center}
%\centering
\begin{tabular}{lll}%\label{table-5}
\toprule
\multicolumn{1}{c}{Parameter} &\multicolumn{1}{c}{MLE} & \multicolumn{1}{c}{AMLE} \\
\midrule
$\alpha$ & 0.983459 & 0.982218\\
$\lambda_1$ & 0.017541 & 0.017622\\
$\lambda_2$ & 0.017541 &0.017622\\
 \bottomrule
\end{tabular} 
\end{center}
\end{table}
\begin{figure}[H]
\begin{center}
\includegraphics[scale=0.6]{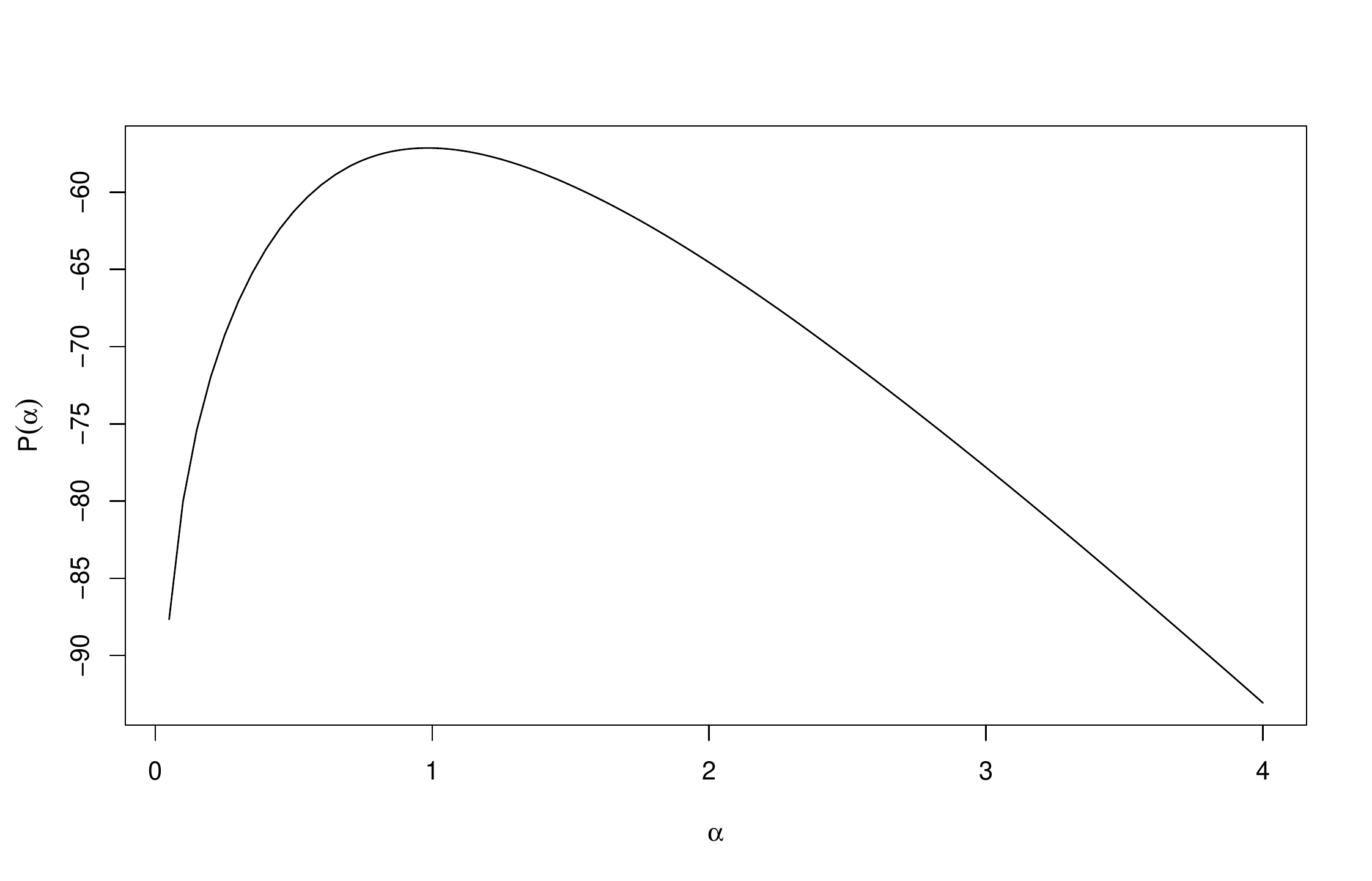}
\caption{profile-loglikelihood function of shape parameter $\alpha$ for scheme-1}
\label{sfig1}
\end{center}
\end{figure}

\begin{table}[H]
\caption{ real data analysis (90\% CI)(scheme-1)}
\label{table-11}
\begin{center}
%\centering
\begin{tabular}{lllll}%\label{table-5}
\toprule
\multicolumn{1}{c}{Parameter} &\multicolumn{2}{c}{ 90\% Asymptotic CI} & \multicolumn{2}{l}{90\% Bootstrap CI} \\
&\multicolumn{1}{l}{LL} & \multicolumn{1}{l}{UL}
&\multicolumn{1}{l}{LL}  &\multicolumn{1}{l}{UL}\\
\midrule
$\alpha$ & 0.6508&1.3160&0.7253&1.5900\\
$\lambda_1$ & 0&0.0426&0.001641&0.05592\\
$\lambda_2$ & 0 & 0.0426&0.001644 & 0.05623\\
 \bottomrule
\end{tabular} 
\end{center}
\end{table}

\noindent {\sc Scheme 2:}
$$
{\bf w} =(1, 3, 4, 5, 5, 13, 14, 31, 44, 51), {\bf z} =(0, 1, 0, 1, 1, 1, 1, 0, 1, 0);
$$

\begin{figure}[H]
\begin{center}
\includegraphics[scale=0.6]{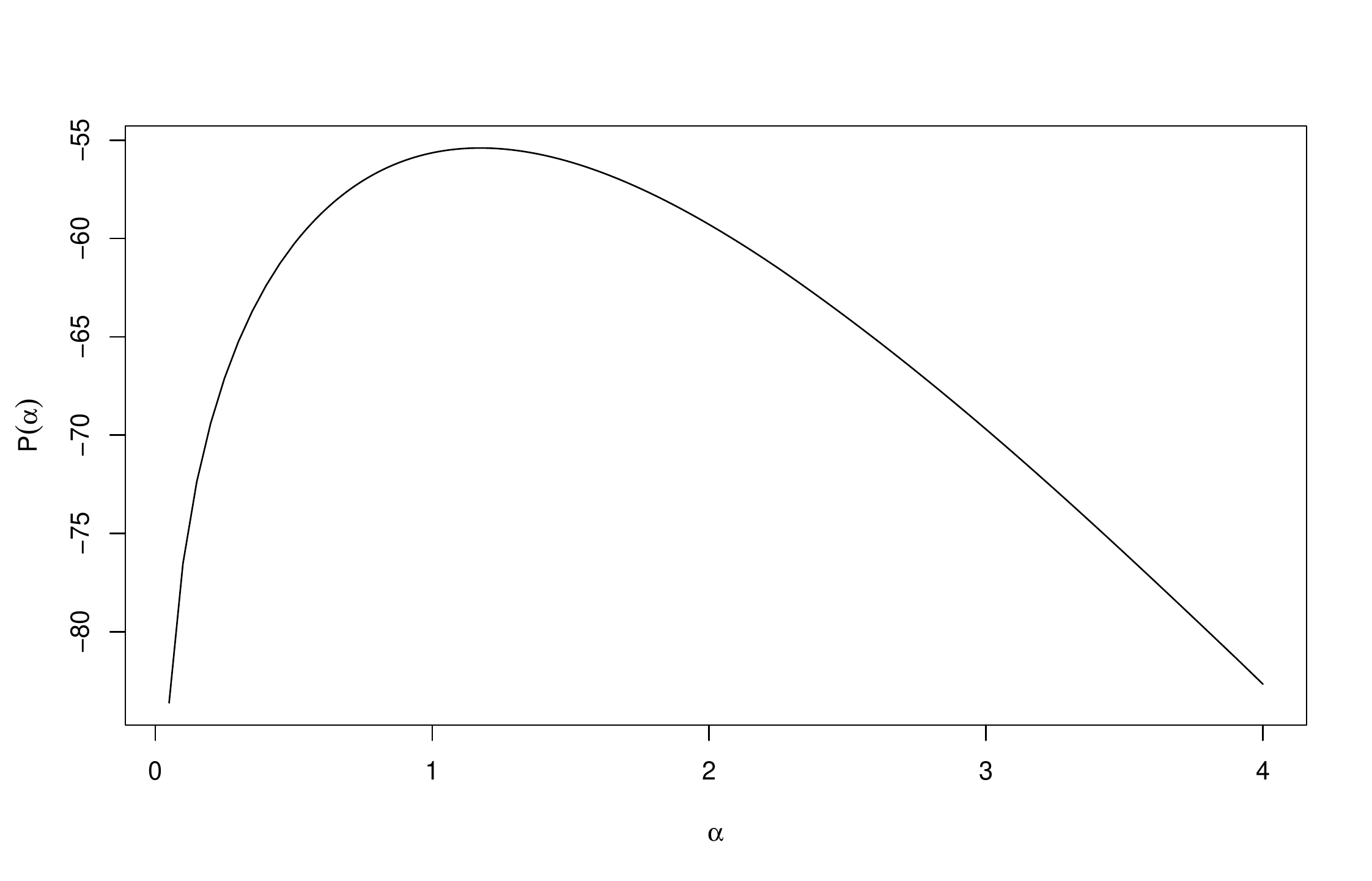}
\caption{profile-loglikelihood function of shape parameter $\alpha$ for scheme-2}
\label{fig-6}
\end{center}
\end{figure}

\begin{figure}[htp]
 \centering
\scalebox{.85}{\begin{tabular}{|c c|}
\hline 
     \includegraphics[width=70mm]{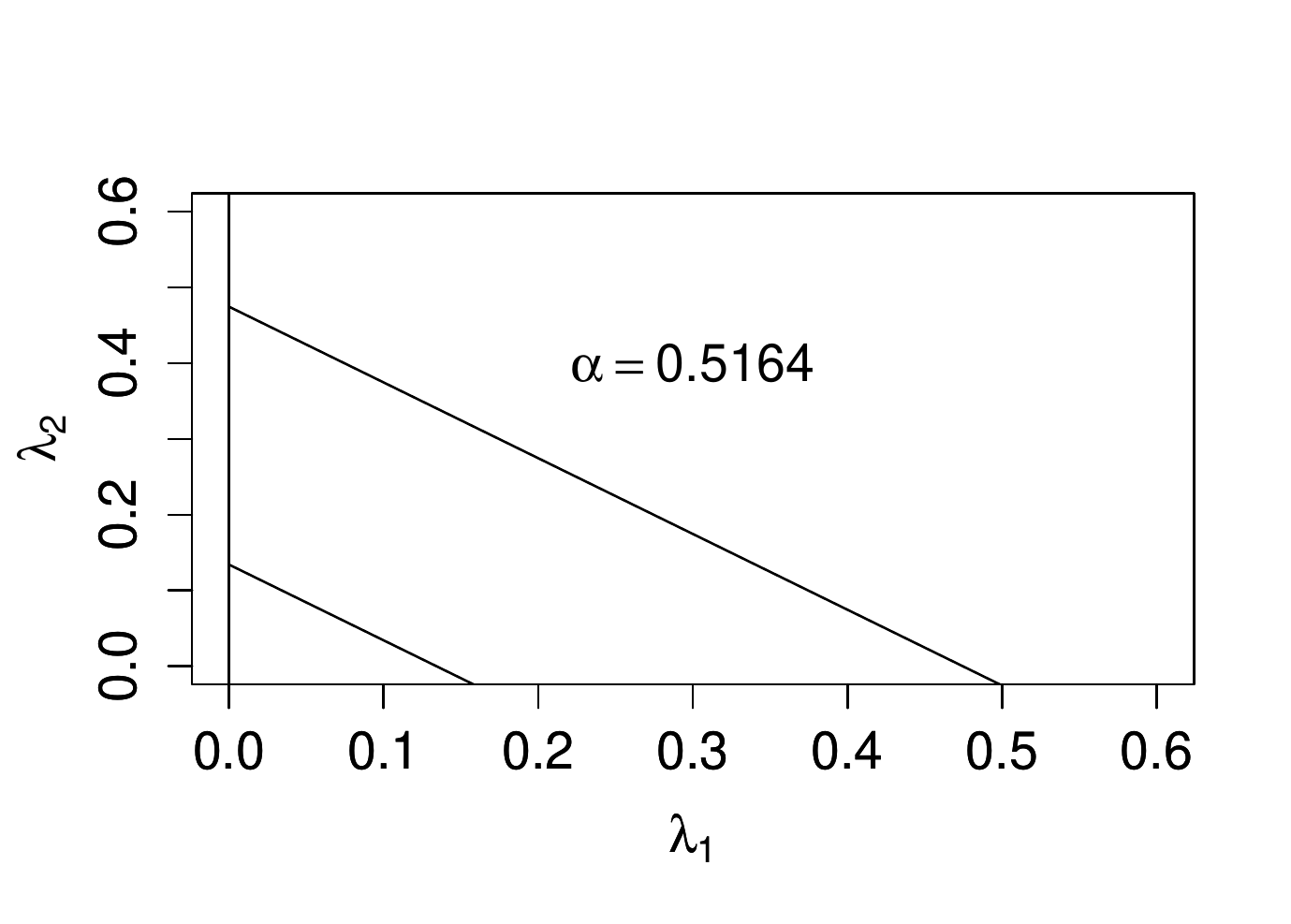}&
     \includegraphics[width=70mm]{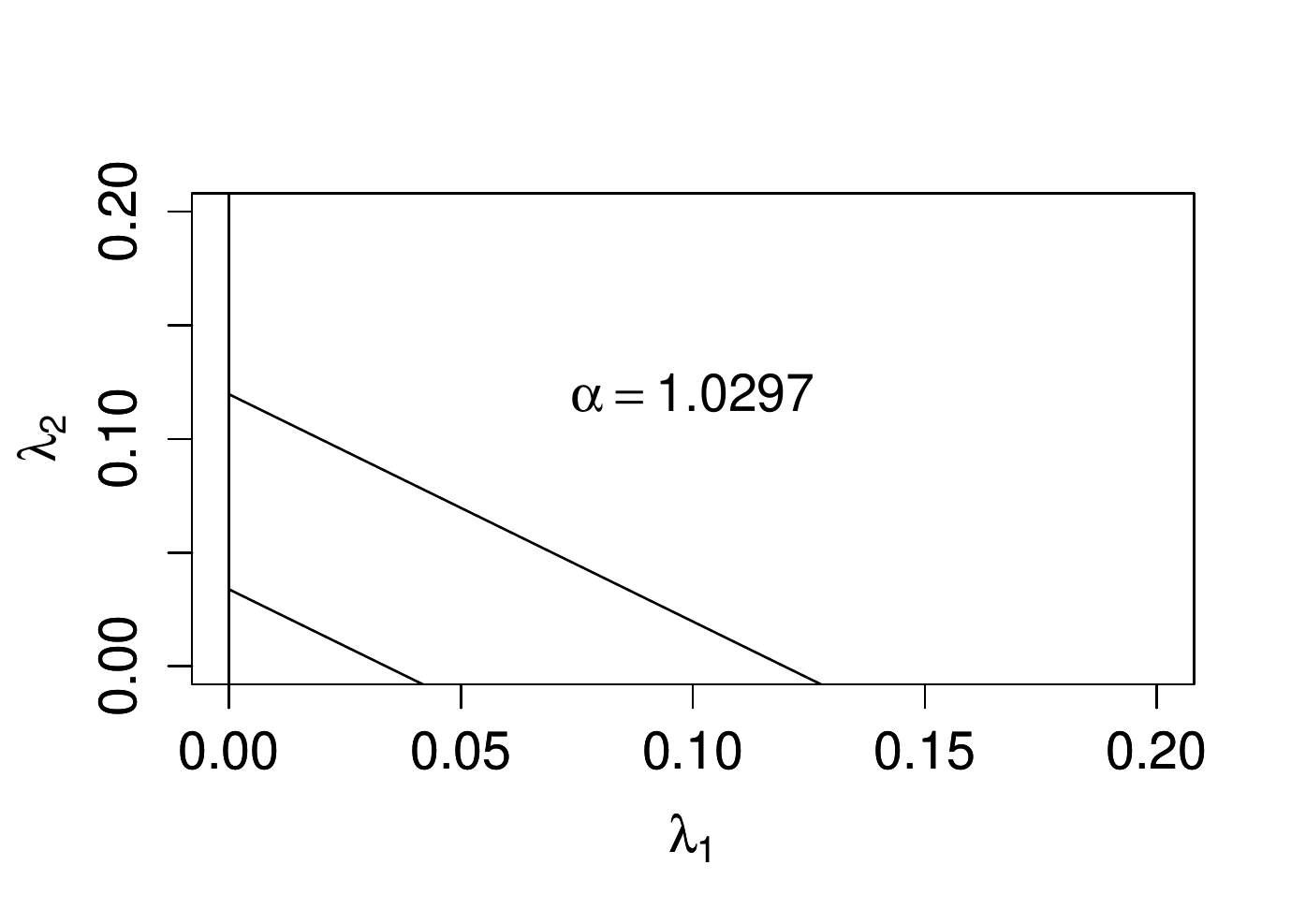}\\
     \includegraphics[width=70mm]{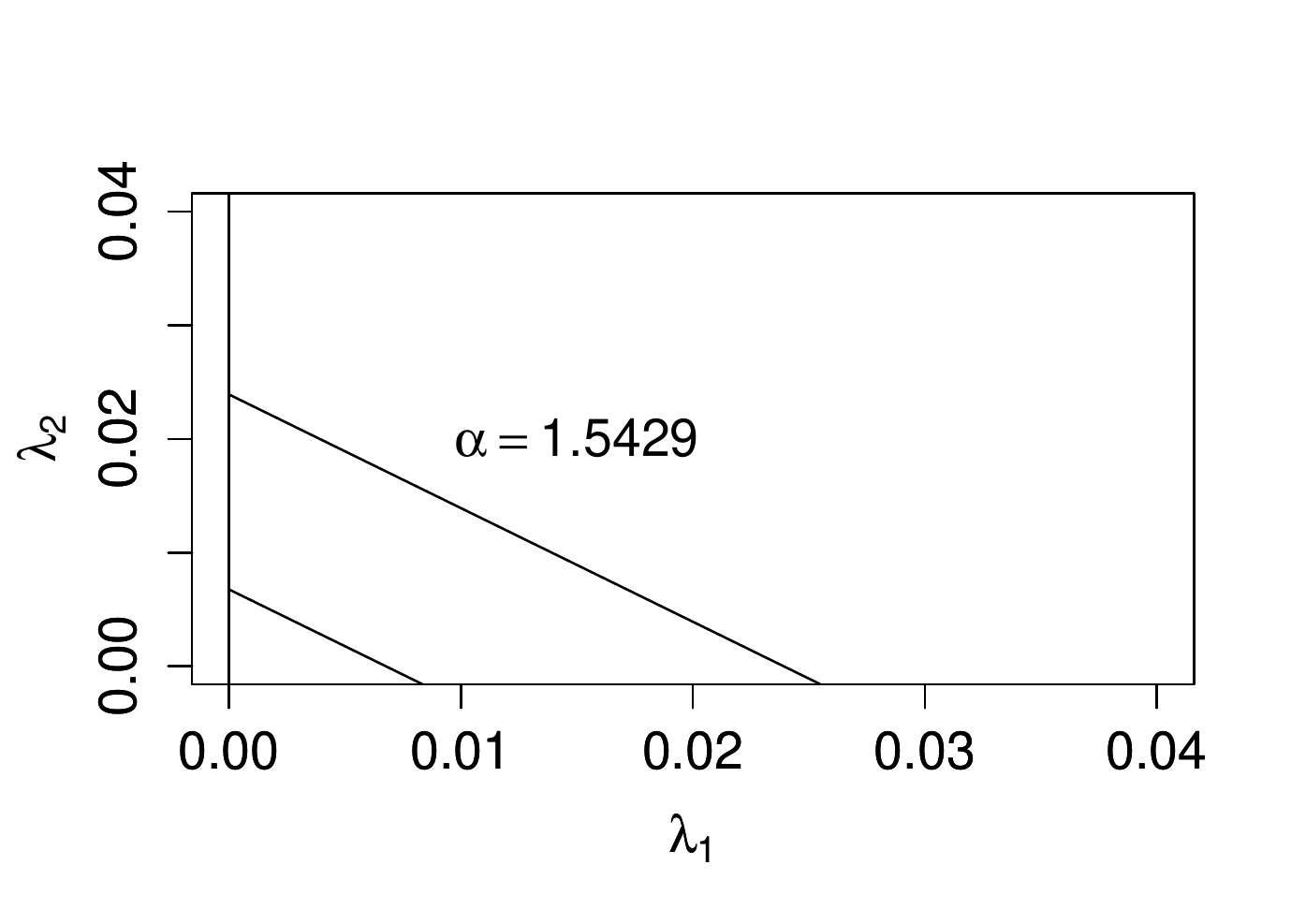}&
     \includegraphics[width=70mm]{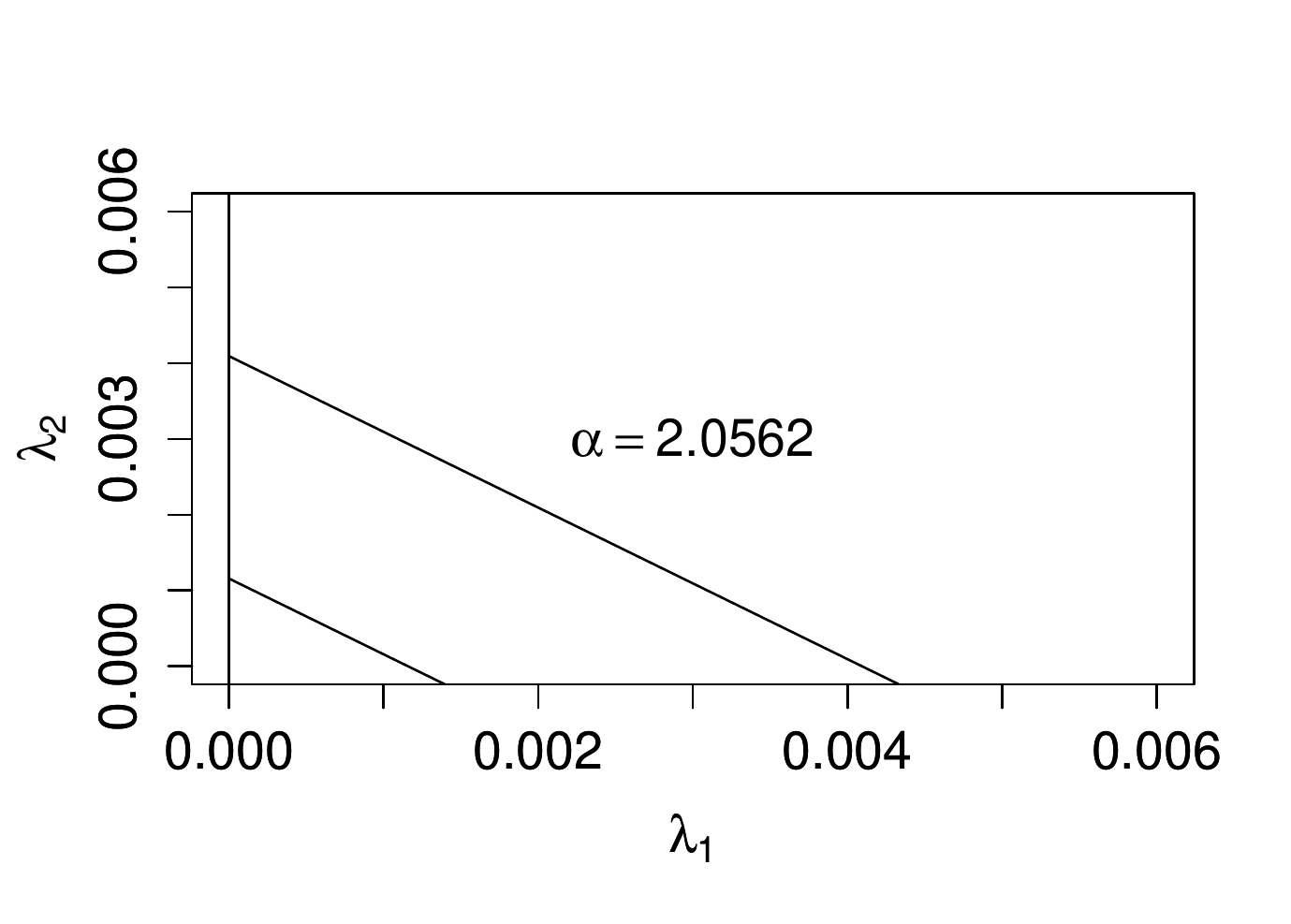}\\
\hline 
\end{tabular}
} 
\caption{Confidence region of $\lambda_1$ and $\lambda_2$ for different values of $\alpha$ for Scheme-2}
\label{fig-8}
\end{figure}

In this case the estimates and the associated confidence intervals are reported in Tables \ref{table-12} and \ref{table-13}.  
The profile log-likelihood function $P(\alpha)$ has been provided in Figure \ref{fig-6} and it indicates that it attains a 
unique maximum.  The confidence set of $(\lambda_1, \lambda_2)$ for different values of $\alpha$ is provided in Figure \ref{fig-8}.

\begin{table}[H]
\caption{ real data analysis(scheme-2)}
\label{table-12}
\begin{center}
%\centering
\begin{tabular}{lll}%\label{table-5}
\toprule
\multicolumn{1}{c}{Parameter} &\multicolumn{1}{c}{MLE} & \multicolumn{1}{c}{AMLE} \\
\midrule
$\alpha$ & 1.1740 & 1.1612\\
$\lambda_1$ & 0.01367 & 0.01421\\
$\lambda_2$ & 0.009116 &0.009479\\
 \bottomrule
\end{tabular} 
\end{center}
\end{table}

\begin{table}[H]
\caption{ real data analysis (90\% CI)(scheme-2)}
\label{table-13}
\begin{center}
%\centering
\begin{tabular}{lllll}%\label{table-5}
\toprule
\multicolumn{1}{c}{Parameter} &\multicolumn{2}{c}{ 90\% Asymptotic CI} & \multicolumn{2}{l}{90\% Bootstrap CI} \\
&\multicolumn{1}{l}{LL} & \multicolumn{1}{l}{UL}
&\multicolumn{1}{l}{LL}  &\multicolumn{1}{l}{UL}\\
\midrule
$\alpha$ & 0.7533&1.5947&0.91046&2.036402\\
$\lambda_1$ & 0&0.03351&0.001241&0.03696\\
$\lambda_2$ & 0 & 0.02303&0.0007122& 0.02532\\
 \bottomrule
\end{tabular} 
\end{center}
\end{table}

\section{\sc Conclusion}
In this paper we analyze the new joint progressive censoring (BJPC) for two populations.  It is assumed that the lifetimes of the 
two populations follow Weibull distribution with the same shape parameter but different scale parameters.  We obtained the MLEs 
of the unknown parameters and since they cannot be obtained in explicit forms we have proposed to use AMLEs which can be obtained
explicitly.  Based on extensive simulation experiments it is observed that the performances of MLEs and AMLEs are very similar 
in nature.  We obtained asymptotic and bootstrap confidence intervals and it is observed that the asymptotic 
confidence intervals perform quite well even for small sample sizes. We further construct an exact joint confidence set of the 
unknown parameters and based on the expected volume of the joint confidence set we have proposed an objective function and it has
been used to obtain optimum censoring scheme.  Note that all the developments in this paper are mainly based on the classical 
approach.  It will be important to develop the necessary Bayesian inference.  It may be mentioned that in this paper we have 
considered the sample sizes to be equal from both the populations,  although most of the results can be extended even when they
are not equal.

\section*{\sc Appendix}
\noindent{\sc proof of Lemma 1:}

$P(\alpha) \rightarrow -\infty  \hspace{5mm} \mbox{as} \hspace{5mm} \alpha \rightarrow 0$ 
\beanno
\lim_{\alpha \to \infty}P(\alpha)&=&-k \lim{\alpha \to \infty}\frac{(\sum_{i=1}^{k-1}(R_i+1)\ln(w_i){w_i}^{\alpha} + (m-\sum_{i=1}^{k-1}(R_i+1))\ln(w_i){w_k}^{\alpha}))}{(\sum_{i=1}^{k-1}(R_i+1){w_i}^{\alpha} + (m-\sum_{i=1}^{k-1}(R_i+1)){w_k}^{\alpha}))} +\sum_{i=1}^{k}\ln(w_i) \\
 &=&-k\ln(w_k)+\sum_{i=1}^{k}\ln(w_i)\\
 &=& \sum_{i=1}^{k}\ln(\frac{w_i}{w_k})<0\\
\eeanno 
$\1 P(\alpha)\rightarrow -\infty \hspace{5mm}as\hspace{5mm}\alpha \rightarrow \infty$
This concludes mle ${\alpha}^*$ is attained in $(0,\infty)$.

According to Balakrishnan and Kateri \cite{balakrishnan2008maximum} $H(\alpha)$ is increasing function of $\alpha$ and $\frac{1}{\alpha}$ is decreasing in $\alpha$ resulting unique solution of equation of equation$(5)$.

\noindent{\sc proof of Lemma 2:}  The proof can be obtained similarly as the proof of Lemma 2 of Mondal and Kundu \cite{MK2016} 
using $m = n$.

\end{document}